\documentclass[a4paper,12pt]{article}
\usepackage{amsmath,amssymb}
\usepackage{graphicx}
\usepackage{blindtext}
\begin{document}
\title{Pion couplings with low-lying nucleon resonances}
\author{Janardan Prasad Singh\footnote{Retired from the services of The Maharaja Sayajirao University of Baroda}  \\ {\scriptsize Physics Department, Faculty of Science, The Maharaja Sayajirao University of Baroda, Vadodara-390002, Gujarat, India}}
\maketitle
\abstract{
We have calculated coupling constants of a neutral pion with the lowest two nucleon resonances. This includes both the diagonal as well as non-diagonal coupling constants involving a nucleon resonance and a nucleon. For this, we first calculate vacuum-to-pion correlation function of the interpolating fields of two nucleons and then take its matrix elements with respect to a nucleon spinor and/or a  nucleon resonance spinor(s). Using different QCD sum rules obtained from different matrix elements we eliminate unwanted coupling constants and solve for the desired ones. This is a simple extension of the projected correlation function approach used in the literature where we use multiple of states involving those of a nucleon and its resonances. We have also checked the stability of our results with respect to variation of different QCD and phenomenological input  parameters.  }\\
\par \hspace{0pt}  \textbf{Keywords :}\textit{ QCD sum rules, Nucleon resonances, non-local operators, Projected correlation function}

\section{Introduction}
Strong interaction is responsible for generating baryonic resonances as well as their couplings with mesons. While the masses and the widths of the baryonic resonances are known fairly well, their coupling constants with mesons are not known so well.
 Determination of these couplings will be useful for construction of NN potential and in other phenomenological analyses such as photoproduction of mesons off a nucleon target. A broad experimental effort has been underway for the past two decades, with measurements of exclusive meson photoproduction and electroproduction reactions, including many polarization observables \cite{burkert}.  Theoretical estimates of these coupling constants have been made using phenomenological Lagrangians \cite{christos,olbrich16,olbrich18} or using QCD based methods. While the phenomenological Lagrangians are good for a broad understanding of hadronic interactions, QCD based approaches give a microscopic view of hadronic interactions and  provide a test of QCD at low energies.   QCD-based methods have been applied, in past, to determine the masses and widths of nucleon resonances \cite{Singh94, kondo06, ohtani}  and other light baryon resonances \cite{Singh07}. In recent past, these methods have also been applied to calculate coupling constant of a pion with a nucleon resonance \cite{kim97, zhu, jido98, azizi, aliev}.  QCD sum rule based on three-point correlation function has been used for calculating $N^{(*)}\pi N^{(*)}$ coupling constant in \cite{azizi}. It is known that an approach based on three-point function gives results that may contain non-negligible contribution from higher resonances $\pi$(1300) and  $\pi$(1800) \cite{aliev}. $N^*N^*\pi$ coupling constant has also been derived in \cite{aliev} within the light-cone QCD sum rule method by considering  vacuum-to-pion correlation function of the interpolating fields of two nucleons. To eliminate the undesired coupling constants involving nucleon and the Roper resonance N(1440), which invariably appears on the phenomenological side, the authors use all the independent Lorentz structures as well as their derivatives including a second derivative for one of them. It is known that the sum rules obtained from derivatives of the correlation functions give bad sum rules\cite{ioffe2010}. Also, though the tensor Dirac structure has nice features for calculating $\pi$NN  coupling constant, other Dirac structures have been found not to be reliable for the same as they contain large contributions from the continuum \cite{kondo03, doi04}.\\
\par \hspace{12pt}   In this work we use a projected correlation function approach as outlined in Ref. \cite{kondo03}. From the vacuum-to pion correlation function of the interpolating fields of two nucleons and by taking its matrix elements with respect to nucleon and/ or nucleon resonance spinor(s), we get different algebraic equations involving desired coupling constants of a pion with a nucleon and nucleon resonances. From these equations  undesired  coupling constants can be eliminated algebraically and desired ones can be extracted.\\
\par \hspace{12pt}   In Sec. 2, we first derive a correlation function of two  interpolating fields of a nucleon between vacuum and one-pion state; next we have derived projected correlation functions from which we get the desired sum rules. In Sec. 3, we first fix on a form of  interpolating field for nucleon and its resonances which is suitable for our purpose. The sum rules constructed from this  interpolating field have been solved    numerically  for different parameter values to get the coupling constants of pions with the two low-lying nucleon resonances. In Sec. 4, we discuss our approach and compare our results  with those available in the literature. Some of the intermediate expressions are given in the Appendix. 

\section{The Sum Rules}
We start with the following correlation function :
\begin{equation} \Pi (q,p)= i \int  \mathrm{d^4}x\,  e^{iqx} \langle 0|  T\{J_p(x) \bar{J_p}(0)\} |\pi^0(p)\rangle   \end{equation}
where $J_p(x)$ is the most general form of the interpolating field of a proton \cite{ioffe2010, doi04} and its resonances \cite{jido96, azizi, aliev} made up of three quark fields without any derivative :\\
\begin{equation} J_p=2\epsilon^{abc} [(u^{aT}C d^b)\gamma_5 u^c+t(u^{aT}C\gamma_5 d^b) u^c] \end{equation}
where \textit{a, b, c} are color indices and t is an arbitrary real parameter. Generally Ioffe current $(t=-1)$\cite{ioffe2010} has been used for cases where only proton is involved. But different values of the parameter t have also  been used \cite{doi04}, particularly in cases where nucleon resonances are involved \cite{jido96, kondo06, azizi, aliev}. Stability of the results with respect to variation in t should be taken as a required condition for checking the suitability of the interpolating field. We parameterize the coupling of $J_p$ with the nucleon and its resonance states as  

\begin{eqnarray} \langle 0|  J_p(0)  |N(k,s)\rangle  =\lambda\, u_N(k,s) \\
      \langle 0|  J_p(0)  |N_+(k,s)\rangle  =\lambda_1\, u_{N_+}(k,s) \\         
    \langle 0|  J_p(0)  |N_-(k,s)\rangle  =\lambda_2\, i\gamma_5\,u_{N_-}(k,s)                                    \end{eqnarray}

where N,  $N_+$  and $N_-$  stand for the nucleon, N(1440) and N(1535) states and + and - denote the parity of the state.   The correlation function $\Pi$(q,p) can be evaluated phenomenologically by saturating it with the nucleon- and its resonance-states. We can also evaluate it using operator product expansion (OPE)  technique exploiting the construction of $J_p$ as given in Eq. (2). For the phenomenological evaluation we use the following Lagrangian \cite{kim97, jido98} :
\begin{equation}
\begin{aligned}
 \mathcal {L}=g_1\bar{N} i  \gamma_5  \overrightarrow{\tau}.\overrightarrow {\pi}  N  + 
  g_2\bar{N_+} i  \gamma_5  \overrightarrow{\tau}.\overrightarrow {\pi}  N_+ + g_3\bar{N_-} i  \gamma_5  \overrightarrow{\tau}.\overrightarrow {\pi}  N_-  +g_4(\bar{N} i  \gamma_5  \overrightarrow{\tau}.\overrightarrow {\pi}  N_+\\
+\bar{N_+} i  \gamma_5  \overrightarrow{\tau}.\overrightarrow {\pi}  N) 
 + g_5(\bar{N}  \overrightarrow{\tau}.\overrightarrow {\pi}  N_-+\bar{N_-} \overrightarrow{\tau}.\overrightarrow {\pi}  N)
+g_6(\bar{N_+}  \overrightarrow{\tau}.\overrightarrow {\pi}  N_- + \bar{N_-}  \overrightarrow{\tau}.\overrightarrow {\pi}  N_+)
\end{aligned}
\end{equation}
The phenomenological expression for the correlation function $\Pi$(q,p) is given in the Appendix. The Wilson coefficients of the OPE are calculated in two steps : the light-cone expansion of the correlation function is performed first. The short-distance  expansion of the vacuum-to-pion matrix elements of light-cone operators is performed next. In the Appendix, we have given the result of parametrization of the vacuum-to pion matrix elements of the light-cone operators.\\
 \par \hspace{12pt}  We define the projected correlation function as \cite{kondo03}
\begin {equation}     \Pi_+^{ij}(q,p) =\bar{u}_i(q,r)\gamma_0\Pi(q,p)\gamma_0u_j(q-p,s)                        \end{equation}  
where each of (i,j) stand for N, $N_+$ or $N_-$. Projected correlation functions have been successfully used for calculating not only diagonal coupling constants, such as $\pi$NN \cite{kondo03} and $\eta$NN, $\eta'$NN \cite{Singh19} coupling constants, but also non-diagonal ones, such as $\pi \Lambda \Sigma$ coupling constant \cite{doi04}.   $\Pi_+^{ij}$(q,p) can be regarded as a function of  $q_0$ in the reference frame in which \textbf{q}=0 . The odd and even parts of $\Pi_+^{ij}(q_0)$ satisfy dispersion relation
\begin{eqnarray}  \Pi_{+o}^{ij}(q_0^2) =-\frac{1}{\pi} \int dq'_0\frac{1}{q_0^2-q^{\prime 2}_0} \textit  {Im} \Pi_+^{ij}(q'_0)\\ 
   \Pi_{+e}^{ij}(q_0^2) =-\frac{1}{\pi} \int dq'_0\frac{q'_0}{q_0^2-q^{\prime 2}_0} \textit{Im} \Pi_+^{ij}(q'_0)      \end{eqnarray}
On taking Borel transform \cite{ioffe2010, kondo03, doi04, Singh19} with respect to $q_0^2$ they take the form
\begin{eqnarray} B[\Pi_{+o}^{ij}(q_0^2)]=\frac{1}{\pi} \int dq'_0\,\  e^{-q^{\prime 2}_0 /M^2} \textit{ Im} \Pi_+^{ij}(q'_0)\\
     B[\Pi_{+e}^{ij}(q_0^2)]=\frac{1}{\pi} \int dq'_0 q'_0\,\  e^{-q^{\prime 2}_0 /M^2}  \textit{Im} \Pi_+^{ij}(q'_0)   \end{eqnarray}
where M is the Borel mass parameter.\\
\par \hspace{12pt} For calculating the correlation function using OPE, we have used vacuum-to-pion matrix elements of non-local operators as given by Eqs. (A2-A5). Following is the  result of our OPE evaluation of the correlation function $\Pi(q,p)$:
\begin{equation}
\begin{aligned}
\Pi(q,p)=-\frac{ t_3}{16\pi^2}i\gamma_5 \left [\frac{\langle \bar{q}q\rangle}{f_\pi}\Big\{(q^2-p.q)ln(-q^2)+ \frac{2}{3}\frac{(p.q)^2}{q^2}\Big\}-18.91\sqrt{2}f_{3\pi}\frac{(p.q)^2}{q^2}\right ]\\
-\frac{t_6}{288}i\gamma_5\frac{\langle \bar{q}q\rangle}{f_\pi}\langle \frac{\alpha_s}{\pi}G^2\rangle \frac{1}{q^2}\left (1+\frac{q.p}{q^2}+\frac{4}{3}\frac{(q.p)^2}{q^4}\right)-i\gamma_5 f_{3\pi}\frac{(q.p)^2}{q^2}\times \\ \left (\frac{t_{11}}{8\sqrt{2}\pi^2}-t_6 \frac{18.91}{72\sqrt{2}} \langle \frac{\alpha_s}{\pi}G^2\rangle \frac{1}{q^4}\right)\\
+i\gamma_5 \not{q}\bigg[\frac{t_1}{72\pi^2}f_{\pi}\delta^2 \frac{q.p}{q^2}(1+\frac{q.p}{q^2})+\frac{t_2}{24\pi^2}f_{\pi}\Big\{-q.p ln(-q^2)+\frac{(q.p)^2}{q^2} \\ -\delta^2\frac{q.p}{q^2}(1+\frac{q.p}{q^2}) +2a'\frac{(q.p)^3}{q^4}\Big\}-\frac{t_4}{9}\frac{(\langle \bar{q}q\rangle)^2}{f_\pi} \frac{q.p}{q^4}(1+\frac{2q.p}{q^2}) \\-\frac{f_\pi}{288}\langle \frac{\alpha_s}{\pi}G^2\rangle\Big\{t_2 \left[\frac{2q.p}{q^4} (1+\frac{2q.p}{q^2})+\frac{4}{3}\delta^2\frac{q.p}{q^6} +24a'\frac{(q.p)^3}{q^8}\right] \\ -\frac{4}{9}t_5 \delta^2\frac{q.p}{q^6}\Big\} -\frac{t_9}{108}\langle \bar{q}g\sigma .Gq\rangle \frac{\langle \bar{q}q\rangle}{f_\pi}\frac{q.p}{q^6}\bigg]\\
+i\gamma_5 \not{p}\bigg[\frac{t_1}{16\pi^2}f_{\pi}\Big\{(q^2-q.p)ln(-q^2)+\delta^2(-ln(-q^2)+\frac{8}{9}\frac{q.p}{q^2})+2a'\frac{(q.p)^2}{q^2}\Big\}\\
+\frac{t_2}{48\pi^2}f_{\pi}\Big\{(2q.p-q^2)ln(-q^2)+\delta^2ln(-q^2)-6a'\frac{(q.p)^2}{q^2}\Big\}\\
-\frac{t_4}{18}\frac{(\langle \bar{q}q\rangle)^2}{f_\pi}\frac{1}{q^2}+\frac{1}{288}\langle \frac{\alpha_s}{\pi}G^2\rangle\Big\{t_5\frac{f_{\pi}}{q^2}\left (1+\frac{q.p}{q^2}-\frac{\delta^2}{9q^2}+4a'\frac{(q.p)^2}{q^4}\right)\\ 
+t_2\frac{f_{\pi}}{q^4}\left(q^2+2q.p+\frac{\delta^2}{3}+12a'\frac{(q.p)^2}{q^2}\right)\Big\}+\frac{t_9}{432}\langle \bar{q}g\sigma .Gq\rangle \frac{\langle \bar{q}q\rangle}{f_\pi}\frac{1}{q^4}\bigg]\\
+\gamma_5 \sigma^{\mu \nu}q_\mu p_\nu\bigg [\frac{-t_3}{48\pi^2} 
\frac{\langle \bar{q}q\rangle}{ f_\pi } \Big\{ln(-q^2)-\frac{q.p}{q^2}-2c'\frac{(p.q)^2}{q^4}\Big\}-
\frac{t_4}{3}\langle \bar{q}q\rangle  f_\pi \frac{1}{q^2}\times\\
\Big\{1+\frac{q.p}{q^2}+\frac{5}{9}\frac{\delta^2}{q^2}(1+\frac{2q.p}{q^2})+4a'\frac{(q.p)^2}{q^4}\Big\}+\frac{t_7}{864}\langle \frac{\alpha_s}{\pi}G^2\rangle \frac{\langle \bar{q}q\rangle}{ f_\pi }\times\\
\frac{1}{q^4}(1+\frac{2q.p}{q^2})+\frac{t_{10}}{18}\langle \bar{q}q\rangle  f_\pi \delta^2 \frac{1}{q^4}
(1+\frac{2q.p}{q^2})-\frac{t_8}{288}\langle \bar{q}g\sigma .Gq\rangle \times\\ \frac{f_\pi}{q^4}
(1+\frac{2q.p}{q^2})\bigg ]
\end{aligned}
\end{equation}
where $t_1=7+6t+7t^2$, $t_2=(1+t)^2$,  $t_3=5+2t-7t^2$,  $t_4=1+2t-3t^2$,  $t_5=19+14t+19t^2$, 
 $t_6=21-6t-15t^2$,  $t_7=17+2t-19t^2$,  $t_8=20+24t-44t^2$,  $t_9=5+6t-11t^2$,  $t_{10}=3-2t- t^2$, $t_{11}=(1-t)^2$. The constants $a'$ and $c'$  are defined as : $a'=3/10 +3a_2/35, c'=3/10+3C_2/35$.
In the above equation, terms such as $\textit{p.q}/q^2$ will be important for $g_4$ and $g_5$ coupling constants. Most of the terms in OPE as given in  Eq. (12) reduce to the corresponding terms given  in Ref.\cite{kondo03} for $t=-1$. However, there are significant  differences  in form of  sign differences of some of the terms, and also  on account  of  higher powers of   $\textit{p.x}$ considered. The sign of $\delta^2$ used in this work is in conformity with that used in Ref. \cite{doi04} and is opposite to that used in Refs. \cite{kondo03} and \cite{kim00} resulting in a sign difference. We can calculate the projected correlation function from the OPE expression as given in Eq. (12) and separate its odd and even parts. It is usual to model the continuum part of $\Pi(q,p)_{ph}$ by  $\Pi(q,p)_{OPE}$ with $q_0^2>s_\pi$, a threshold parameter and only terms with postive powers of $q_0$ are taken into account in doing so \cite{Singh94, Singh07, ioffe2010, Singh19, kondo03, doi04}. The parameters $\delta^2$ and $C_2$ are defined as \cite{novikov, belyaev, agaev11}  $$\langle 0| \bar{q}g\tilde{G}_{\mu \nu}\gamma^\nu q |\pi^0(p)\rangle =i p_\mu f_\pi \delta^2$$  $$C_2=-\frac{f_\pi f_{3\pi}}{\langle\bar{q}q\rangle}\frac{6.44}{\sqrt{2}}$$ In Eq. (12),  $a_2$ and  $C_2$ are the coefficients of the Gegenbauer polynomials in the leading twist-2 pion wave function and two-particle pion wave function of twist-3 respectively\cite{belyaev}. $f_{3\pi}$ appears in two-particle wave function of twist-3 as well as in three-particle pion wave function \cite{belyaev}. It has been estimated that  $a_2\simeq 0.4$ and $f_{3\pi}\simeq 0.0045 GeV^2$. The coefficient of $\bar{u}_N i \gamma_5 u_N$ in the phenomenological as well as in the OPE expressions of the projected correlation function  have been Borel transformed and are given below : 
\begin{equation}
\begin{aligned}
 B[\Pi_{+o}^{NN}]_{ph}=\lambda^2 g_1(m) e^{-\frac{m^2}{M^2}} (\frac{2m}{M^2}-\frac{g'_1}{g_1}+\frac{1}{2E_{k1}}+\frac{1}{2m})+\lambda_1^2 g_2(m_1) e^{-\frac{m_1^2}{M^2}}\times\\
\frac{ (m+m_1)(3m-m_1-2\omega_{p1})}{4m_1 E_{k2}}(\frac{2m_1}{M^2}-\frac{g'_2}{g_2}+\frac{1}{2E_{k2}}+\frac{1}{2m_1})\\
+\lambda_2^2 g_3(m_2) e^{-\frac{m_2^2}{M^2}}
\frac{ (m_2-m)(3m+m_2-2\omega_{p1})}{4m_2 E_{k3}}(\frac{2m_2}{M^2}-\frac{g'_3}{g_3}+\frac{1}{2E_{k3}}+\frac{1}{2m_2})\\
+\lambda\lambda_1 g_4(m_1) e^{-\frac{m_1^2}{M^2}}
\frac{ (m+m_1)(m-\omega_{p1})}{2m_1 E_{k4}}(\frac{2m_1}{M^2}-\frac{g'_4}{g_4}+\frac{1}{2E_{k4}}+\frac{1}{2m_1})\\
+\lambda\lambda_1 g_4(m) e^{-\frac{m^2}{M^2}}
\frac{ (3m-m_1-2\omega_{p1})}{2 E_{k'4}}(\frac{2m}{M^2}-\frac{g'_4}{g_4}+\frac{1}{2E_{k'4}}+\frac{1}{2m})\\
+\lambda\lambda_2 g_5(m_2) e^{-\frac{m_2^2}{M^2}}
\frac{ (m_2-m)(m-\omega_{p1})}{2m_2 E_{k5}}(\frac{2m_2}{M^2}-\frac{g'_5}{g_5}+\frac{1}{2E_{k5}}+\frac{1}{2m_2})\\
-\lambda\lambda_2 g_5(m) e^{-\frac{m^2}{M^2}}
\frac{ (3m+m_2-\omega_{p1})}{2 E_{k'5}}(\frac{2m}{M^2}-\frac{g'_5}{g_5}+\frac{1}{2E_{k'5}}+\frac{1}{2m})\\
+\lambda_1\lambda_2 g_6(m_2) e^{-\frac{m_2^2}{M^2}}
\frac{ (m_2-m)(3m-m_1-2\omega_{p1})}{4m_2E_{k6}}(\frac{2m_2}{M^2}-\frac{g'_6}{g_6}+\frac{1}{2E_{k6}}+\frac{1}{2m_2})\\
-\lambda_1\lambda_2 g_6(m_1) e^{-\frac{m_1^2}{M^2}}
\frac{(m_1+m) (3m+m_2-2\omega_{p1})}{4m_1 E_{k'6}}(\frac{2m_1}{M^2}-\frac{g'_6}{g_6}+\frac{1}{2E_{k'6}}+\frac{1}{2m_1})\\
\end{aligned}
 \end{equation}

\begin{equation}
\begin{aligned}
B[\Pi_{+o}^{NN}]_{OPE}= M^2\Big [-t_1\frac{f_{\pi}}{8\pi^2}L^{-4/9}E_{k1}\omega_{p1}+t_2\frac{f_{\pi}}{12\pi^2}L^{-4/9}E_{k1}\omega_{p1}  \\ +\frac{t_3}{48\pi^2} \frac{\langle \bar{q}q\rangle }{f_{\pi}} (m+E_{k1}-3\omega_{p1}) \Big ]  E_0\left(\frac{s_\pi}{M^2}\right) \\  +\frac{t_1}{9\pi^2}f_{\pi}\delta^2 E_{k1}\omega_{p1}+\frac{t_4}{3}\langle \bar{q}q\rangle f_\pi (m+ E_{k1})\\
\frac{1}{M^2}\Big[(-t_4\frac{5}{27}+\frac{t_{10}}{18})\langle \bar{q}q\rangle f_{\pi}\delta^2(m+E_{k1})L^{32/75}-\frac{1}{288} \langle \frac{\alpha_s}{\pi}G^2\rangle \\ \times \Big\{2t_5 f_{\pi} L^{-4/9}E_{k1}\omega_{p1} +4t_2f_{\pi}L^{-4/9}E_{k1}\omega_{p1} +t_6 \frac{\langle \bar{q}q\rangle}{f_{\pi}}\omega_{p1} \\ -\frac{t_7}{3} \frac{\langle \bar{q}q\rangle}{f_{\pi}}(m+E_{k1}) \Big\} 
-\frac{t_9}{72}f_{\pi}\langle \bar{q}g\sigma .Gq\rangle L^{-14/27}(m+E_{k1})\Big]
\end{aligned}
\end{equation}
\begin{figure}[h]
\centering
\includegraphics[width=0.3\linewidth]{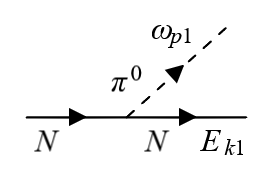}
\includegraphics[width=0.3\linewidth]{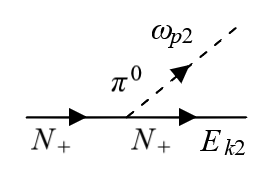}
\includegraphics[width=0.3\linewidth]{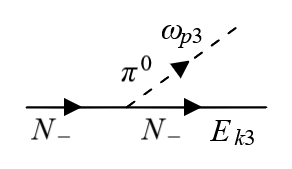}\\
[\baselineskip]
\includegraphics[width=0.3\linewidth]{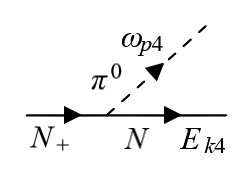}
\includegraphics[width=0.3\linewidth]{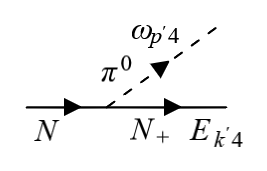}
\includegraphics[width=0.3\linewidth]{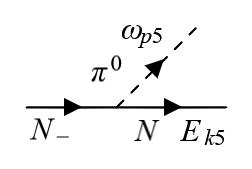}\\
[\baselineskip]
\includegraphics[width=0.3\linewidth]{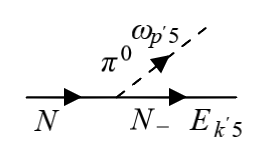}
\includegraphics[width=0.3\linewidth]{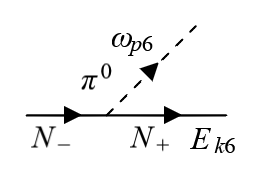}
\includegraphics[width=0.3\linewidth]{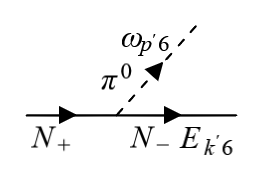} 
\caption{Feynman diagrams for the phenomenological form of the correlation function}
\label{fig:feyn}
\end{figure}
In above equations $\textit{m}$, $m_1$ and $m_2$ are masses of N, $N_+$ and $N_-$ states. 
The continuum contribution of B[$\Pi_{+o}^{NN}]_{ph}$, modelled as explained above, has been transferred and combined with B[$\Pi_{+o}^{NN}]_{OPE}$ in form of $E_0$, where $E_0(x)=1-e^{-x}$.
 $g'_i$ denotes derivative of the respective coupling constant.  Though $\omega_{p1}\simeq$0.01 GeV is small, it has been retained for the sake of demonstration. 
   We have done a similar exercise  with $\Pi_{+o}^{N_+N_+}$, $\Pi_{+o}^{N_-N_-}$,  $\Pi_{+o}^{N_+N}$ and $\Pi_{+o}^{N_-N}$ also getting five equations in all. L= $\alpha_s(\mu^2)/\alpha_s(M^2)$ and it appears in the OPE expressions of the correlation function for renormalization group improvement; the anomalous dimensions of various operators have been taken from  Refs. \cite{ioffe2010, agaev11}. The emitted pion has significant energies, $\omega_{p4}\simeq$0.42 GeV for $N_+\to N$  and  $\omega_{p5}\simeq$0.49 GeV for  $N_-\to N$. 
A characteristic feature of $(\Pi^{N_-j}_{+o})_{ph}$ is that due to appearance of $\gamma_5$ in the matrix element  $\langle0|J_p(0)|N_-(k,s)\rangle$ the pole at $q^2= m^2_2$ gets removed. We are giving below the expression for  B[$\Pi_{+o}^{N_-N}]_{OPE}$ with higher powers of  $\omega_{p5}$ to show its characteristic difference.
\begin{equation}
\begin{aligned}
 B[\Pi_{+o}^{N_-N}]_{OPE}=M^2\Big[\left(\frac{t_2}{24\pi^2}- \frac{t_1}{16\pi^2}\right)f_{\pi}L^{-4/9}(E_{k5}+m-\omega_{p5})\omega_{p5} +\frac{t_3}{48\pi^2}\frac{\langle\bar{q}q\rangle}{f_{\pi}} \\  \times (E_{k5}+m-3\omega_{p5})\Big]   E_0\left(\frac{s_{\pi}}{M^2}\right) \\
+\frac{t_1}{72\pi^2}f_{\pi}\delta^2[4(E_{k5}+m)\omega_{p5}-5\omega_{p5}^2] +\frac{t_2}{24\pi^2}f_{\pi}\delta^2\omega_{p5}^2-\frac{t_3}{24\pi^2}\frac{\langle\bar{q}q\rangle}{f_{\pi}}c' \\ \times (E_{k5}+m)\omega_{p5}^2+\frac{t_4}{3}f_{\pi}\langle\bar{q}q\rangle (E_{k5}+m) \\
+\frac{1}{M^2}\Big[\frac{t_4}{3}\langle\bar{q}q\rangle\Big\{-f_{\pi} (E_{k5}+m)(\frac{5}{9}\delta^2L^{32/75}+4a'\omega_{p5}^2)-\frac{2}{3}\frac{\langle\bar{q}q\rangle}{f_{\pi}}L^{4/9}\omega_{p5}^2\Big\} \\ +\frac{t_{10}}{18}\langle\bar{q}q\rangle f_{\pi}\delta^2L^{32/75}  (E_{k5}+m) -\frac{1}{288}\langle \frac{\alpha_s}{\pi}G^2\rangle \Big\{t_5 f_{\pi} L^{-4/9} \\ \times (E_{k5}+m-\omega_{p5})\omega_{p5}   +2t_2f_{\pi}L^{-4/9}(E_{k5}+m+\omega_{p5})\omega_{p5}  \\ +t_6\frac{\langle\bar{q}q\rangle}{f_{\pi}}\omega_{p5} -\frac{t_7}{3}\frac{\langle\bar{q}q\rangle}{f_{\pi}}(E_{k5}+m)\Big\}  -\frac{t_9}{72}f_{\pi}\langle \bar{q}g\sigma .Gq\rangle L^{-14/27}(E_{k5}+m) \Big]
\end{aligned}
\end{equation}

\begin{figure}[h]
\centering
\includegraphics[width=0.45\linewidth]{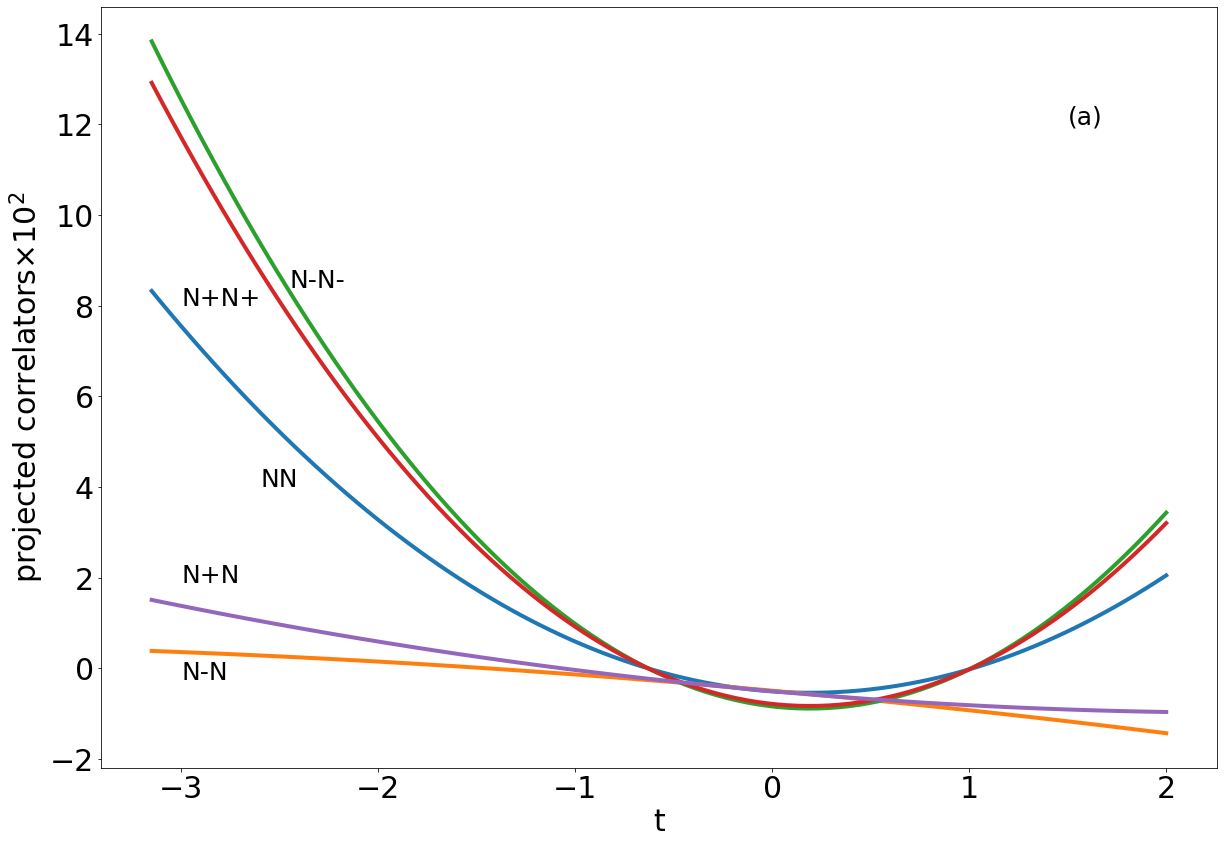}
\includegraphics[width=0.45\linewidth]{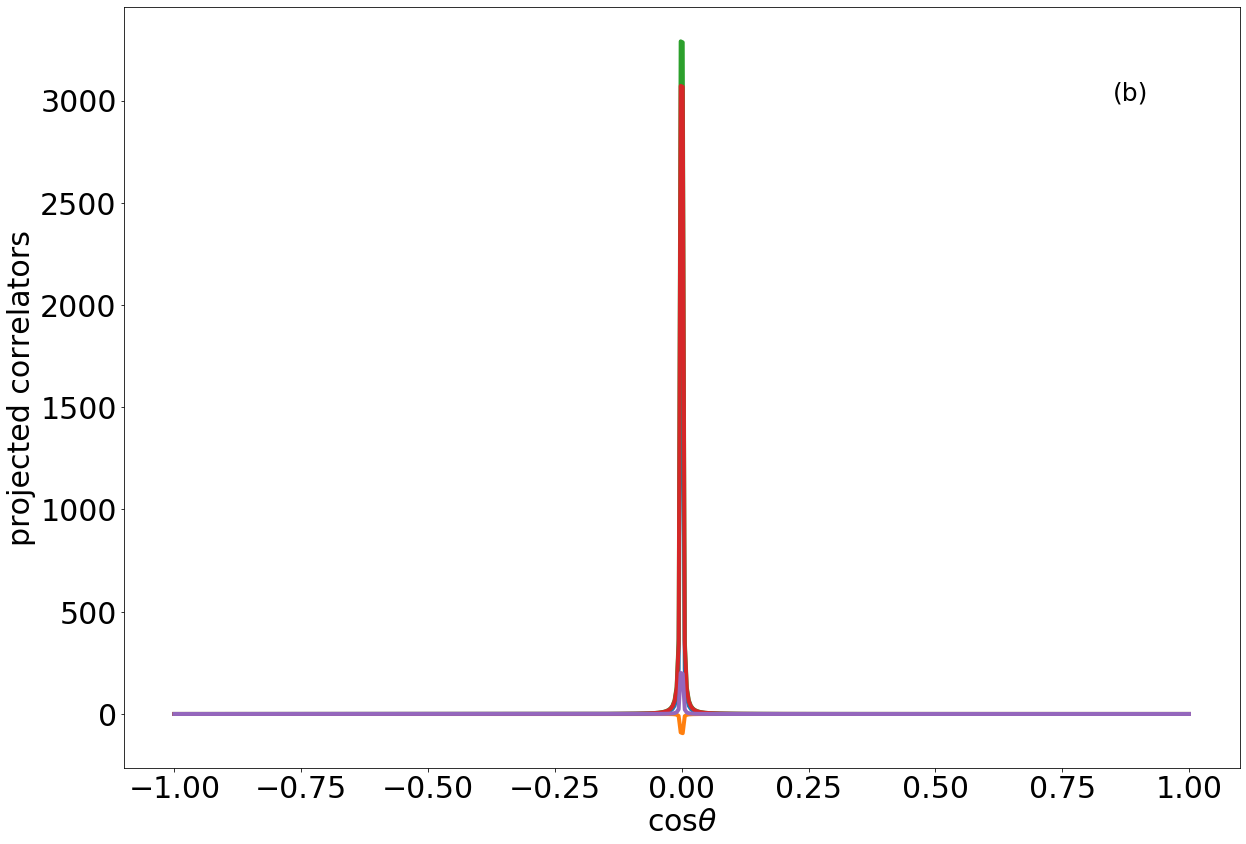}
\caption{{\scriptsize  Plots of $B[\Pi_{+o}^{NN}]$, $B[\Pi_{+o}^{N_+N_+}]$, $B[\Pi_{+o}^{N_-N_-}]$, $B[\Pi_{+o}^{N_+N}]$ and  $B[\Pi_{+o}^{N_-N}]$, obtained from OPE, as a function of t for a typical case with  $M^2$= 3.0 $GeV^2$, $s_{\pi}$=2.7 $GeV^2$, $\langle\bar{q}q\rangle=-0.0117$ $GeV^3$ and other QCD parameters as given in  Table 2 have been shown in Fig. (a). Plots of the same functions as functions of $cos\theta$ ($t=tan\theta$) have been shown in Fig. (b).}}
\label{fig:proj_t}
\end{figure}
We have shown plots of $B[\Pi_{+o}^{NN}]$, $B[\Pi_{+o}^{N_+N_+}]$, $B[\Pi_{+o}^{N_-N_-}]$, $B[\Pi_{+o}^{N_+N}]$ and  $B[\Pi_{+o}^{N_-N}]$, obtained using OPE, as a function of $t$ in Fig. 2a and as a function of $cos\theta$ in Fig. 2b, where $t=tan\theta$,  for a typical case with  $M^2$= 3.0 $GeV^2$,  $s_\pi$=2.7 $GeV^2$,  $\langle\bar{q}q\rangle=-0.0117$ $GeV^3$ and other QCD parameters as given in  Table 2. Fig. 2a shows behavior of the function close to the origin whereas Fig. 2b shows the behavior for large $t$.  We observe stability of the diagonal projected correlation functions $B[\Pi_{+o}^{NN}]$, $B[\Pi_{+o}^{N_+N_+}]$ and $B[\Pi_{+o}^{N_-N_-}]$ at $t\simeq 0.2$, similar to an observation made  in \cite{doi04}. Qualitatively similar behavior is obtained for $M^2$= 2.5 $GeV^2$, 3.5 $GeV^2$ and $\langle\bar{q}q\rangle$=$-0.0145$ $GeV^3$. 
In the next section we will consider a combination of all the five projected correlation functions to extract the desired coupling constants of pions with the nucleon and its resonances; there we will test the stability of the combination of  projected correlation functions with respect to variation in t in a more comprehensive way.
\section{Numerical Estimates}
We have nine coupling constants of a neutral pion with N, $N_+$ and $N_-$ states including the three non-diagonal ones with different arguments. The $\pi$NN coupling constant $g_1$ can be determined by retaining only the first term in Eq. (13) and a lower continuum threshold
 $s'_\pi \simeq 2.0 GeV^2$ on  the OPE side as is done in a standard way \cite{kim00, kondo03, doi04}. Using four equations obtained from the four projected correlation functions we can eliminate four coupling constants, namely, $g_4$(\textit{m}), $g_5$(\textit{m}), $g_6(m_1)$ and $g_6(m_2)$ getting an equation involving coupling constants $g_2(m_1)$, $g_3(m_2)$, $g_4(m_1)$  and $g_5(m_2)$ (hereafter referred to simply as $g_2$, $g_3$, $g_4$ and $g_5$).  As is clear from Eq. (13), we also get  $g'_i$ as an unknown constant along with  each  $g_i$ on the phenomenological side of the sum rule. We have checked the value of  $g'_1$ from the sum rule for  $g_1$ only for different values of  $\langle\bar{q}q\rangle$ and for different values of $M^2$ in the range where the sum rule is applicable. We found that  $-1.0$ $ GeV^{-1}<g'_1/g_1<0.55$  $GeV^{-1}$. We assume that $g'_i/g_i$ (i=2,3,4,5) also lies in approximately the same range.
 Calling $g'_i/g_i=\alpha$  for i=2,3 and $g'_i/g_i=\beta$ for i=4,5, we will vary  $\alpha$ and $\beta$  independently between -1.0 Ge$V^{-1}$ and 0.55 Ge$V^{-1}$ and check the resulting variations in $g_i$'s. The equation obtained from combining different projected correlation functions, as stated above, can be solved for $g_2$, $g_3$, $g_4$  and $g_5$ in such a way that the phenomenological  and the OPE sides  of the combination of the  projected correlation functions match over a range of Borel mass parameter which covers the continuum threshold. \\
\par \hspace{12pt} Before we proceed further, we will check the stability of our results with respect to variation of the parameter t. In Figs. (3a, 3b) we have shown plots of a combination of projected correlation functions containing coupling constants $g_2$,  $g_3$, $g_4$ and $g_5 $. While Fig. (3a) displays the behavior of the function close to the origin and shows 

\begin{figure}[h]
\centering
\includegraphics[width=0.45\linewidth]{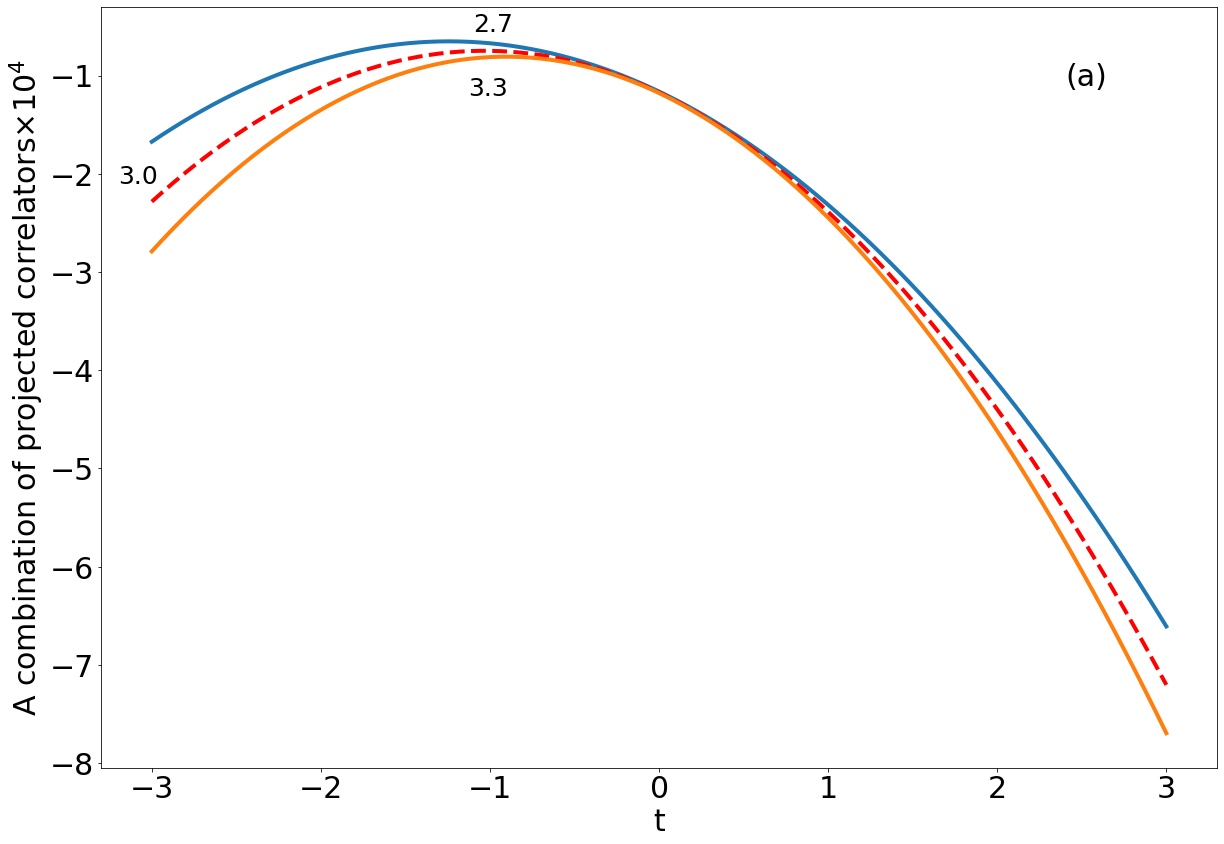}
\includegraphics[width=0.45\linewidth]{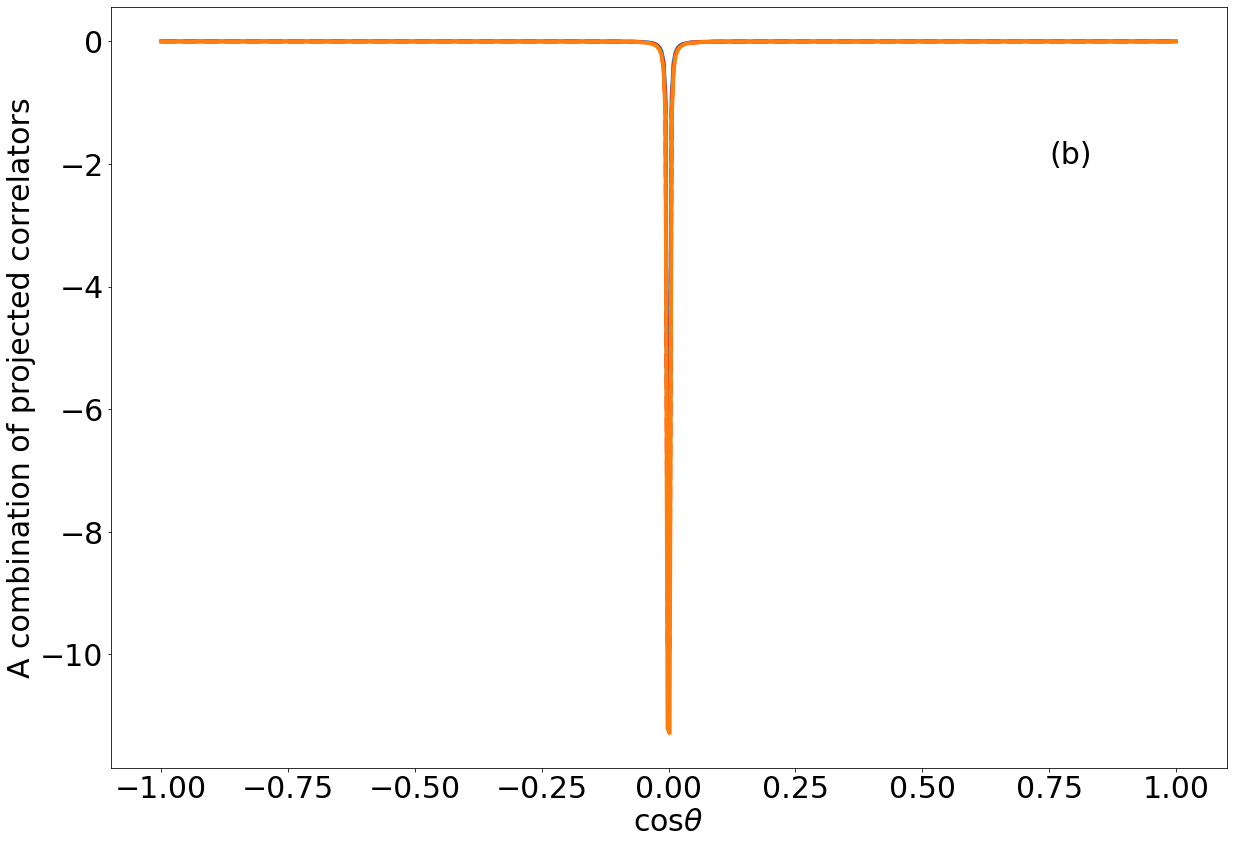}
\caption{{\scriptsize  Plots of a combination of projected correlation functions as a function of $t$ for $ s_\pi=2.7$ $GeV^2$, $\langle\bar{q}q\rangle=-0.0117$ $GeV^3$, other QCD parameters as given in  Table 2  and $M^2$=2.7$GeV^2$, 3.0 $GeV^2$ and 3.3 $GeV^2$ have been shown in Fig. (a). Plots of the same combination of projected correlation functions as  functions of $cos\theta$ ($t=tan\theta$) have been shown in Fig. (b).}}
\label{fig:mass_t}
\end{figure}
 a maxima at $t\simeq-1.0$, in Fig (3b) the same function has been plotted with respect to $cos \theta$ ($t=tan\theta$) for displaying its behavior for large t. We will also determine the coupling constants $\lambda(t), \lambda_1(t)$ and $\lambda_2(t)$ as given in Eqs (3-5) from a simultaneous fit of mass sum rule for N, $N_+$ and $N_-$ ignoring the mass widths of $N_+$ and $N_-$; an explicit demonstration will be given later. The Wilson coefficients with t-dependence and including radiative corrections have been taken from \cite{ohtani} and consistancy with the results given in \cite{gruber} for $t=-1$ was checked. For the evaluation we have used $\alpha_s$ to two-loop order including $\Lambda_{QCD}$=0.355 GeV from \cite{yndurain}. We have plotted chirality conserving part of the 
\begin{figure}[h]
\centering
\includegraphics[width=0.45\linewidth]{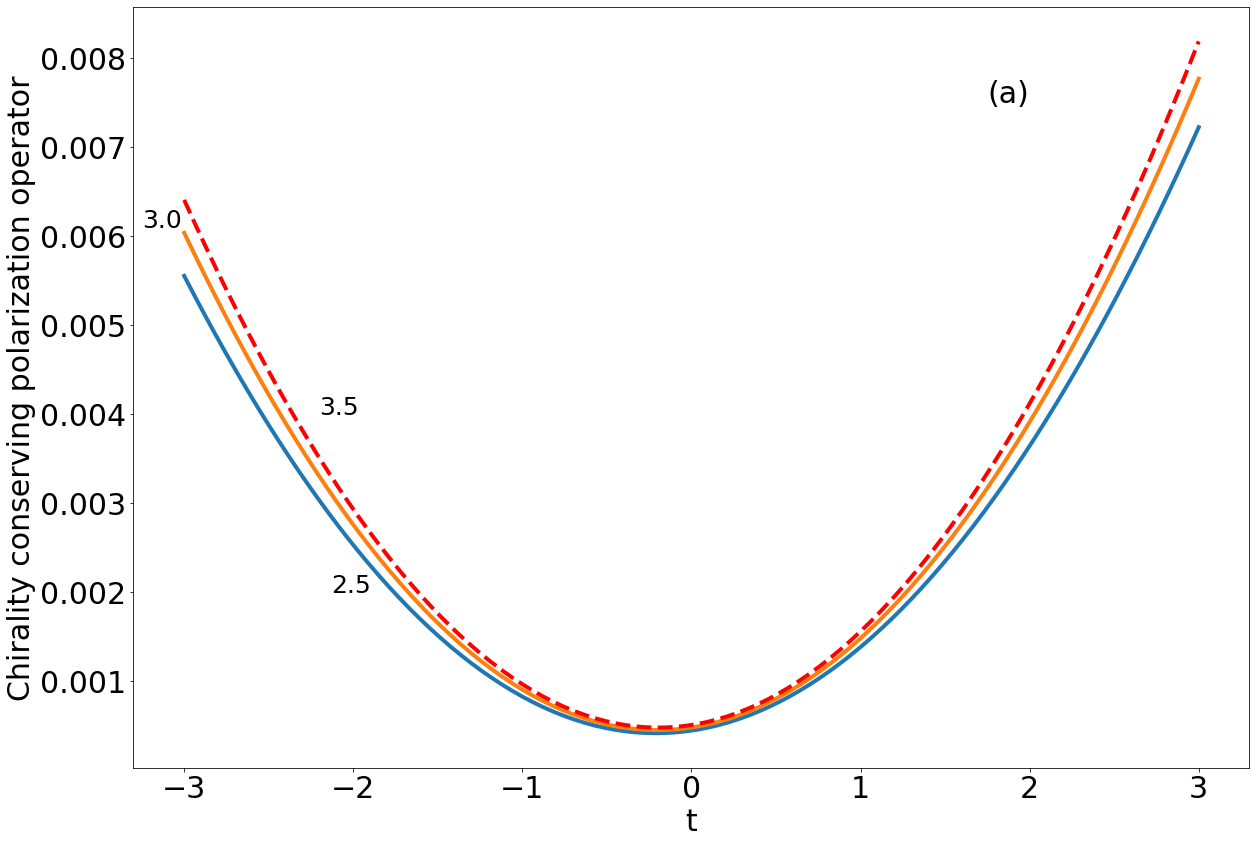}
\includegraphics[width=0.45\linewidth]{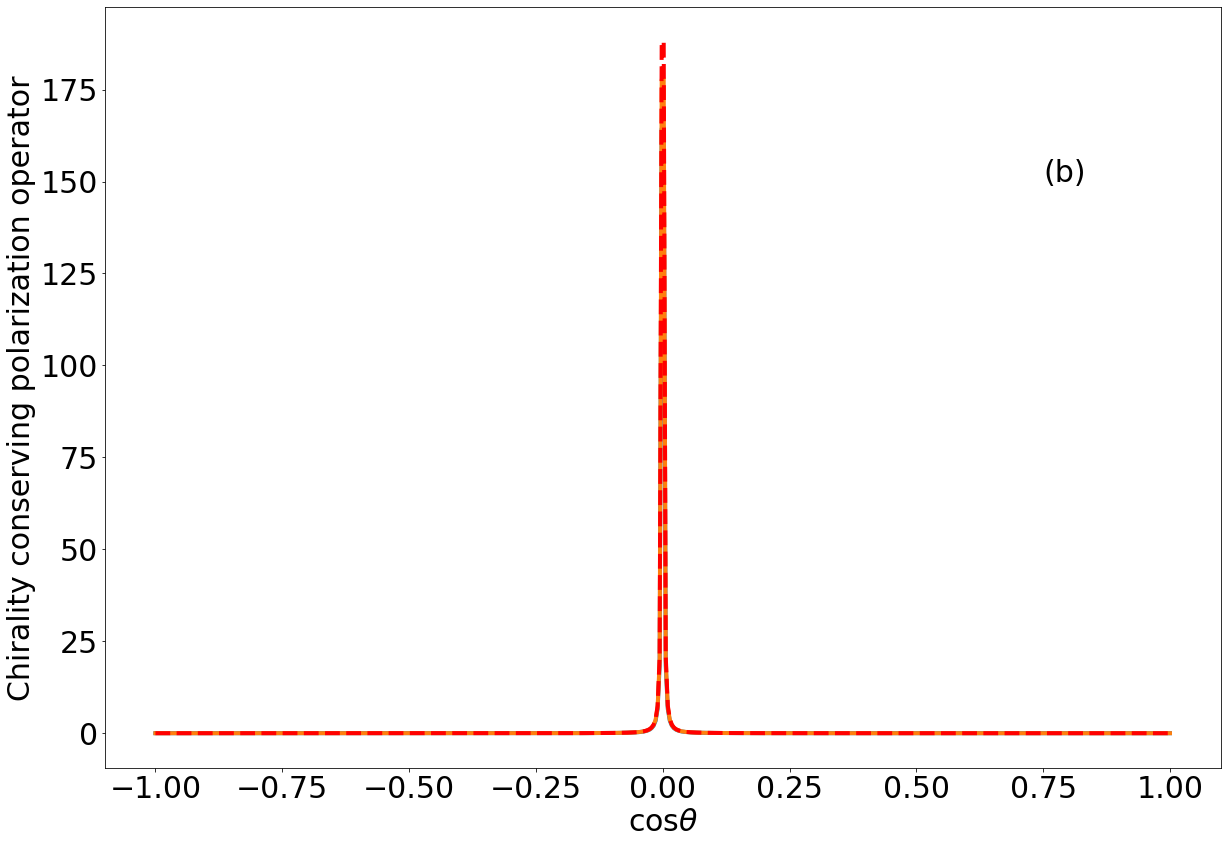}
\caption{{\scriptsize  Plots of the chirality conserving part of the polarisation operator constructed from the interpolating field given in Eq. (2) as functions of  $t$  for $ s_0=2.7$ $GeV^2$, $\langle\bar{q}q\rangle=-0.0117$ $GeV^3$, other QCD parameters as given in  Table 2  and $M^2$=2.5 $GeV^2$, 3.0 $GeV^2$ and 3.5 $GeV^2$ have been shown in Fig. (a). Plots of the same functions as  functions of $cos\theta$ ($t=tan\theta$) have been shown in Fig. (b). The three lowest mass states N,  $N_+$ and $N_-$ are explicitly included in the physical spectrum. }}
\label{fig:combn_t}
\end{figure}
polarization operator, constructed from the interpolating field $J_p$ given in Eq. (2), as a function of t in Fig. (4a); the same function has been plotted as a function of cos$\theta$ in Fig. (4b) to display its behavior for large t. The plot shows a minimum at $t\simeq-0.2$ and is qualitatively similar to the one given in Ref. \cite{doi00} where a plot of chirality non-conserving part of polarization operator without radiative correction is given. It is found that the ratio of the combination of projected correlation functions to the chirality conserving part of the polarization operator has a plateau for $t \lesssim -2.0$. Based on these observations, we have determined the coupling constants $g_i$'s in three regions of t: close to the extremums of the two curves, namely around $t=-0.2$ and around $t=-1.0$, and around $t=-2.5$ where the ratio of the two functions has a plateau. At each value of t, we first determine $\lambda(t), \lambda_1(t)$ and $\lambda_2(t)$ which were, in turn, used to determine  $g_i(t)$ for i=2, 3, 4. In Table 1, we have displayed our results of $g_i(t)$'s assuming $g_5$ is known from \cite{an11, olbrich18}. The numerical values of  $g_i(t)$ close to $t=-0.2$ are unstable and unreasonably large. The numerical values of $g_i(t)$ at $t=-2.5\pm0.5$ are relatively stable but values of $g_4$ are unreasonably large, and hence we discard both these results. In contrast, the numerical values of $g_i(t)$'s around  $t=-1$ are reasonable and the results are stable over a range $t=-1.0\pm0.1$. In view of these observations, from here onwards we will take $t=-1.0$ and analyze the results in more detail.\\

\begin{table}[h!]
\caption{{\footnotesize Our results on coupling constants of a neutral pion with low-lying nucleon resonances for  various parameter values, $\langle \bar{q}q\rangle$ =$-0.0117$ Ge$V^3$ and other QCD parameters as given in Table 2}}
\begin {tabular}{l l l l c l l l l}
\hline \hline
$ t $  &  $s_\pi$  &  $\alpha$  &  $\beta$  &  {\scriptsize range of $M^2$ where phen. and OPE sides match}    &  $g_2$  &   $g_3$  &   $g_4$  &   $g_5$\\
\hline \hline
-0.1 & 2.7 & -1.0  &  -1.0  &  2.8 - 3.2  &  12.2  &  -9.62  & 350.0  &  -1.1\\
-0.2 & 2.7 & -1.0  &  -1.0  &  2.8 - 3.35  & 213.1  &  -6.34  & 188.3  &  -1.1\\
-0.3 & 2.7 & -1.0  &  -1.0  &  2.8 - 3.25  &162.0  &  -4.5  & 180.0  &  -1.1\\
\hline
-0.8 & 2.7 & -1.0  &  -1.0  &  2.25 - 2.75  &8.0  &  7.0  & 19.5  &  -1.1\\
-0.9 & 2.7 & -1.0  &  0.0  &  2.25 - 2.7  &6.0  &  14.1  & -7.5  &  -1.1\\
-0.95 & 3.0 & 0.0  &  0.0  &  2.4 - 3.0  &7.65  &  12.6  & -9.2  &  -1.1\\
-1.0 & 2.7 & -1.0  &  -1.0  &  2.6- 3.15  &6.0  &  13.1  & -10.4  &  -1.1\\
-1.0 & 3.0 & 0.0  &  -1.0  &  2.65 -3.25  &7.0  &  14.5  & -8.8  &  -1.1\\
-1.05 & 2.7 & -1.0  &  -1.0  &  2.7 -3.35  &7.4  &  11.5  & -17.3  &  -1.1\\
-1.1 & 2.7 & 0.0  &  -1.0  &  2.8- 3.5  &5.0  &  6.8  & -18.0  &  -1.1\\
-1.3 & 2.7 & -1.0  &  -1.0  &  2.65 - 3.2  &-28.6  &  11.4  & -28.0  &  -1.1\\
\hline
-2.0 & 2.7 & -1.0  &  -1.0  &  2.6 - 3.0  &15.4  &  10.6 & -95.0  &  -1.1\\
-2.5 & 2.7 & -1.0  &  -1.0  &  2.7 - 3.15  &13.0  &  10.9 & -89.0  &  -1.1\\
-3.0 & 2.7 & -1.0  &  -1.0  &  2.75 - 3.4  &10.9  &  14.3 & -80.0  &  -1.1\\
\hline
\end{tabular}
\label{table:1}
\end{table}

  In Fig. 5, we have shown a fit for the chirality conserving part of the polarization operator constructed from the interpolating field given in Eq. (2) $(t=-1)$ when the physical spectrum is saturated with the three lowest mass states over a range of $M^2=(1.7 - 3.3) GeV^2$. This is a demonstration of a typical case for   $\langle \bar{q}q\rangle=-0.0117$ Ge$V^3$, where the three  coupling constants appearing in Eqs. (3), (4) and (5) are uniquely fixed as $\lambda^2=7.256\times10^{-4}$ $GeV^6$,    $\lambda_1^2=6.027\times10^{-4}$ $GeV^6$ and $\lambda_2^2=1.369\times10^{-4}$ $GeV^6$.  A similar fit for $\langle \bar{q}q\rangle=-0.0145$ Ge$V^3$ yields  $\lambda^2=7.758\times10^{-4}$ $GeV^6$,    $\lambda_1^2=6.242\times10^{-4}$ $GeV^6$ and $\lambda_2^2=1.210\times10^{-4}$ $GeV^6$. \\

\par \hspace{12pt}
\begin{figure}[h]
\centering
\includegraphics[width=0.8\linewidth]{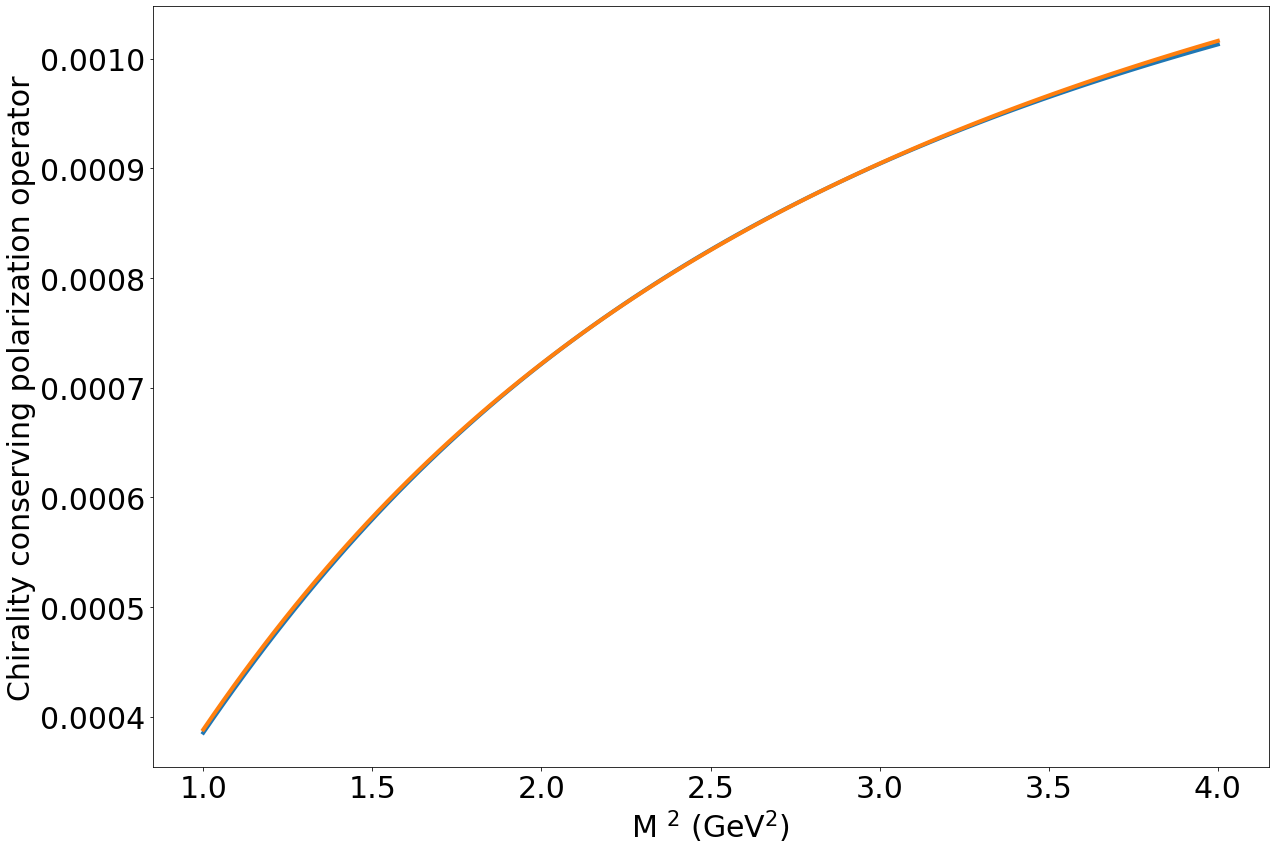}
\caption{{\scriptsize Plots of the chirality conserving part of the polarisation operator constructed from the interpolating field given in Eq. (2) $(t=-1)$ as a function of Borel mass squared $M^2$. The three lowest mass states N,  $N_+$ and $N_-$ are explicitly included in the physical spectrum. }}
\label{fig:mass}
\end{figure}

\par \hspace{12pt} It may be emphasized that in the current approach separating the contributions of $g_1$ and $g_2$ is not difficult even though N and $N_+$ states have the same parities. As is clear from Eq. (13), the dominant behaviors of the coefficients of  $g_1, g_2$ and $g_3$ in the considered range of $M^2$ are determined by the exponential factors. The parities of the states in the projected correlation function approach have a minor role in the  form of an overall multiplicative factor.
\begin{figure}[h]
\centering
\includegraphics[width=0.45\linewidth]{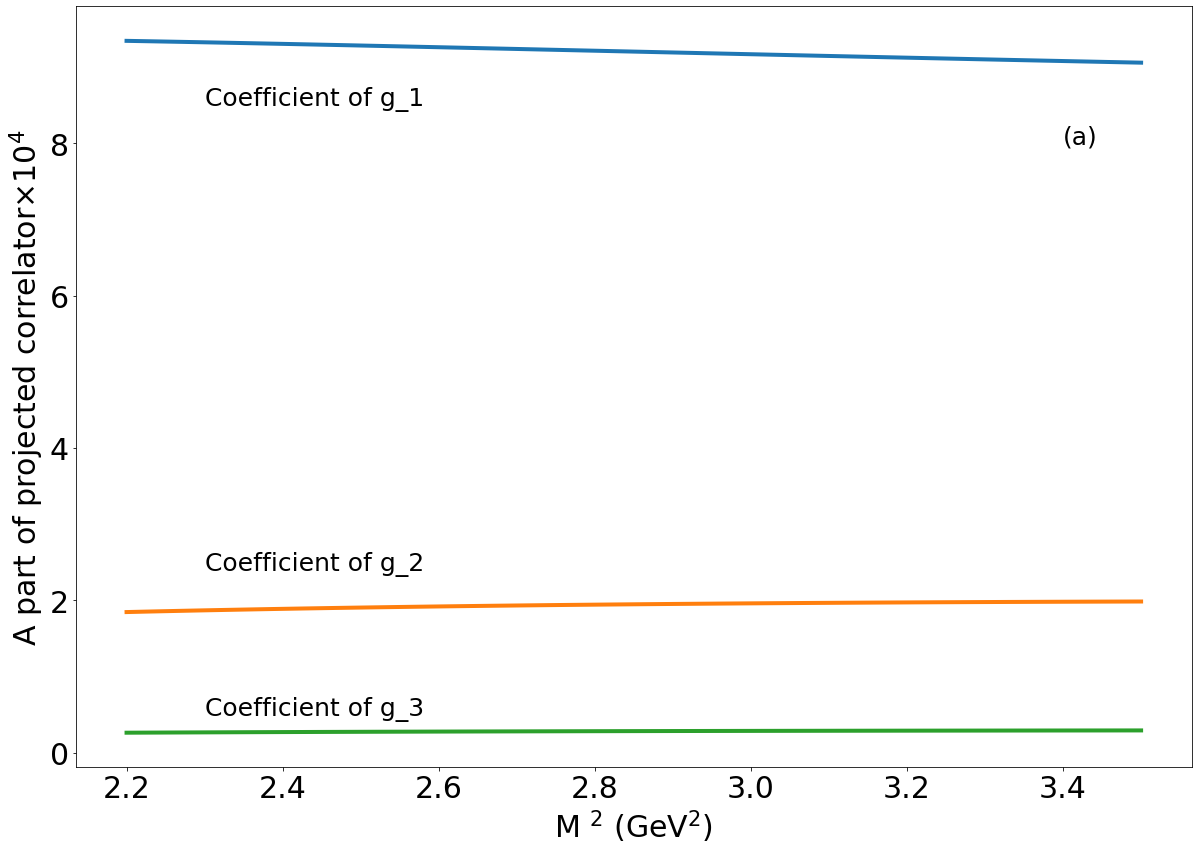}
\includegraphics[width=0.45\linewidth]{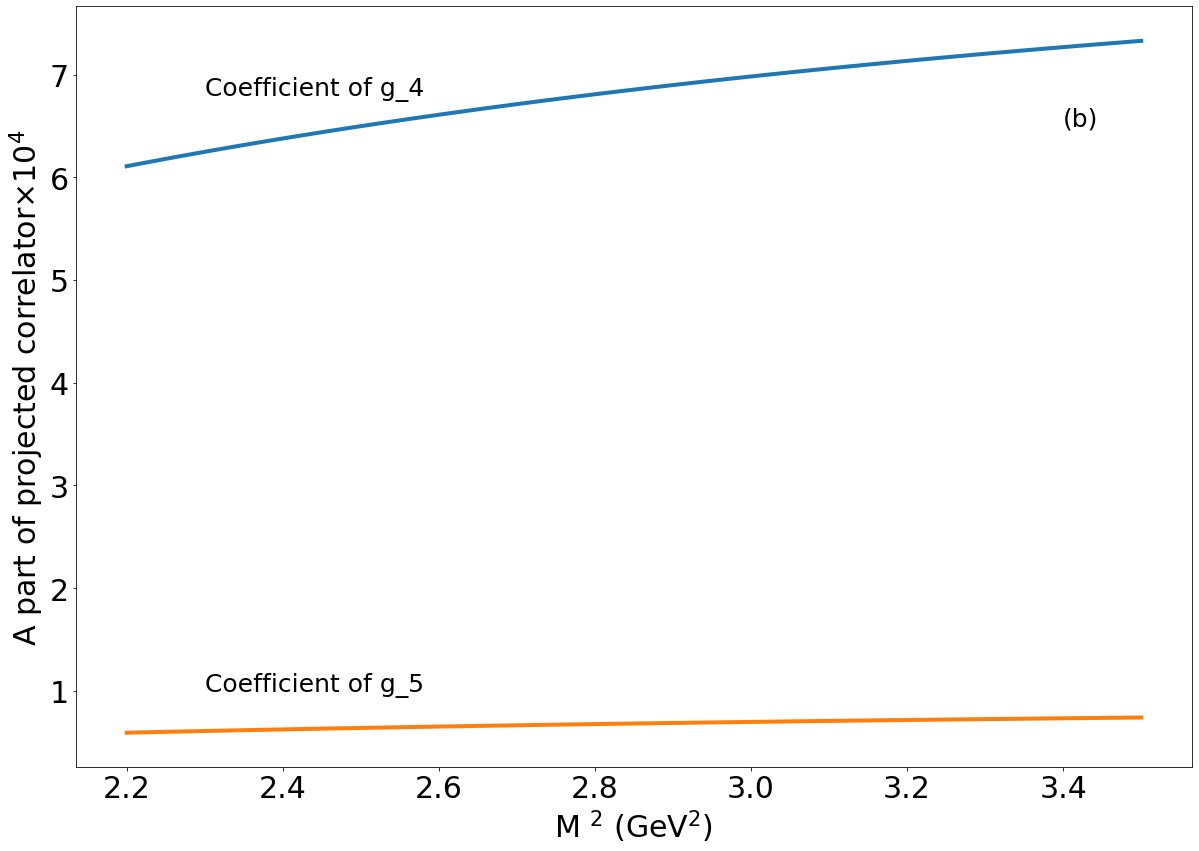}
\caption{{\scriptsize Plots of the coefficients of $g_1, g_2$ and $g_3$ as a function of Borel mass squared $M^2$ as they appear in Eq. (13) have been shown in Fig.(a). Plots of the coefficients of $g_4$ and $g_5$ as a function of Borel mass squared $M^2$ as they appear in Eq. (13) have been shown in Fig.(b). }}
\label{fig:fig2}
\end{figure}
 We have shown plots of coefficients of $g_1, g_2$ and $g_3$ appearing in Eq. (13) in Fig. (6a) and of coefficients  of $g_4$ and $g_5$  in Fig. (6b). It is clear that in addition to different exponential behaviors there is a wide gap between the coefficients of $g_1$ and $g_2$ which is the result of large gap between the two masses \textit{m} and $m_1$. Similar behavior is observed for other B[$\Pi_{+o}^{ij}]_{ph}$. The gap between the coefficients of  $g_2$ and $g_3$ is not so large, though they have opposite parities, largely because the masses of $N_+$ and $N_-$ states are so close. \\
 \par \hspace{12pt}   For determining $g_2$,  $g_3$  and $g_4$ from our sum rules, we have also assumed $-1.1\leq g_5\leq-0.6$ based on estimates given in Ref.\cite{an11,olbrich18}  and references therein.  We have  varied  the continuum threshold  $s_\pi$ between  (2.7 - 3.0) Ge$V^2$ which covers the mass of the  next nucleon resonance N(1650) along with its width. For matching the two sides of the combination of the projected correlation functions we have chosen the  range of $M^2$ to lie between  (2.7 - 3.3) Ge$V^2$. The reason for this choice is  as follows : As stated earlier, the Wilson coefficients of OPE are based on light-cone expansion as well as on short-distance  expansion. For light-cone  expansion it has been argued \cite{aliev}that the upper limit is obtained by requiring that higher states and continuum contributions constitute, say, 40\% of the perturbative contribution. The lower limit of $M^2$  is determined from the condition that higher twist contributions are less than the leading twist contributions. Our chosen range of $M^2$ is consistent with this criterion. We will comment later on this choice from the point of view of the short distance expansion.   In the literature, \cite{ioffe2010, Singh07, Singh19, kondo03, doi04} a range for the  values of the quark condensate  $ \langle \bar{q}q\rangle$ has been suggested, hence we have varied it between  $-(0.0117 - 0.0145)$ Ge$V^3$ and solved the equation. In Table 3, we have  listed a sample of our results obtained from variation of various parameters. In Figs. (7- 10) we have shown plots of  some of our results for the sake of illustration. It is observed that the largest changes in the values of the coupling constants occur due to changes in   $ \langle \bar{q}q\rangle$ which appears as a factor in most of the terms of  of the OPE expression  for  $\Pi$(q,p). Furthermore, we note  that  it is  $g_4$ which undergoes maximum change. Changes in the coupling constants due to variations in $\langle \frac{\alpha_s}{\pi}G^2\rangle$, $m_0^2$ and  $\delta^2$ within the range  suggested in the literatureare, is small and hence  we have ignored them. The mass sum rule of nucleon and its resonances gives only a magnitude of $\lambda$,  $\lambda_1$ and  $\lambda_2$;  we have chosen signs of  $\lambda, \lambda_1$ and $\lambda_2$ to be +ve. If we reverse the signs of $\lambda$ or $\lambda_1$ and  $\lambda_2$, then the signs of $g_4$ and $g_5$ will get reversed.\\
\par \hspace{12pt}  We could have done a similar analysis with $\Pi_{+e}^{ij}$. However, this sum rule has a larger dependence  on the continuum threshold  $s_\pi$ \cite{kondo03, doi04} resulting  in a  larger uncertainty on our end reults of $g_i$'s. Hence, we have not tried  it. \\
\par \hspace{12pt} We have replaced the first term  containing  $g_1$ in  B$[\Pi_{+o}^{ij}]_{ph}$ with the OPE expression with lower continuum threshold  $s'_\pi$  as mentioned earlier\cite{kim00, kondo03, doi04}. Based on the current literature \cite{ioffe2010, kondo03,  doi04, Singh19,  kim00, belyaev}, in Table 2 we have listed the numerical values of the QCD and phenomenological parameters used in this work. For  $\langle \bar{q}q\rangle$ and $s_\pi$ a range of their values is given.

\begin{table}[h!]
\begin{center}
\caption{{\footnotesize Numerical values of the QCD and phenomenological parameters \\ used  in this work ($\langle \bar{q}g\sigma .Gq\rangle$= $m_0^2\langle \bar{q}q\rangle$) }} \label{table:1}
\begin {tabular}{ lc lc lc lc lc lc l }
\hline \hline
 $\Lambda_{QCD}$ &  $\langle \bar{q}q\rangle$ & $\langle \frac{\alpha_s}{\pi}G^2\rangle $ &  $m_0^2$ & $\delta^2$ & $f_\pi$    \\
 (GeV) & ($GeV^3$) & ($GeV^4$) & ($GeV^2$) & ($GeV^2$) & (GeV)   \\
\hline
0.355 &-0.0117 & 0.012 & 0.8 & 0.2 & 0.093 \\
   &-0.0145 &   &   &   & \\  
\hline \hline
$f_{3\pi}$  &  $s_\pi$ &  $s'_\pi$ &    $m_1$  &  $m_2$ & $a_2$ \\
($GeV^2$) & ($GeV^2$) & ($GeV^2$)  & (GeV) & (GeV) &   \\
\hline
0.0045 & 2.7 & 2.0 &  1.44 & 1.54 & 0.4 \\
   & 3.0 &  &   &   &  \\
\hline
\end{tabular}
\end{center}
\end{table}

 We are giving below  a typical equation involving  the coupling constants we want to determine for   typical  parameter values   $\langle \bar{q}q\rangle$=$-0.0117$ Ge$V^3$,   $s_\pi$= 2.7 Ge$V^2$, $\alpha$=$\beta$ =$-1.0$ and the remaining parameters as given in Table 2 at $M^2$=2.8 Ge$V^2$:
\begin {equation}0.917 g_2 +6.263 g_3 +1.238 g_4 +3.641 g_5=69.977\end{equation}
Had we used the phenomenological expression for the first term containing $g_1$ in B$[\Pi_{+o}^{ij}]_{ph}$ with the experimental value $g_1$=13.0 in place of the OPE expression as done above, we would have got 70.553 on the rhs in Eq. (16) making little difference on the end results. We have displayed in Table 3, a sample of results for the coupling constants obtained along with the parameters used and the range of $M^2$  over which the phenomenological and the OPE sides of the expressions match. We have also shown some of these results graphically.\\
\begin{table}[h!]
\caption{{\footnotesize Our results on coupling constants of a neutral pion with low-lying nucleon resonances for various parameter values}}
\begin {tabular}{l l l l c l l l l}
\hline
$\langle \bar{q}q\rangle$  &  $s_\pi$  &  $\alpha$  &  $\beta$  &  {\scriptsize range of $M^2$ where phen. and OPE sides match}    &  $g_2$  &   $g_3$  &   $g_4$  &   $g_5$\\
\hline
-0.0117 & 2.7 & -1.0  &  -1.0  &  2.6 - 3.15  &  6.0  &  13.1  &  -10.4  &  -1.1\\
-0.0117 & 3.0 & -1.0  &  -1.0  &  2.6 - 3.2  &  6.9  &  14.1  &  -10.5  &  -0.6\\
-0.0117 & 3.0 & 0.0  &  0.0  &  2.7 - 3.25  &  7.8  &  13.2  &  -10.4  &  -1.1\\
-0.0117 & 2.7 & 0.0  &  0.0  &  2.7 - 3.25  &  7.9  &  11.48  &  -10.4  &  -0.6\\
-0.0117 & 2.7 & -1.0  &  0.0  &  2.7 - 3.25  & 7.0  &  11.25  &  -11.2  &  -0.6\\
-0.0117 & 3.0 & -1.0  &  0.0  &  2.7 - 3.25  &  7.0  &  12.95  &  -11.2  &  -1.1\\                 
-0.0117 & 3.0 & 0.0  &  -1.0  &  2.65 - 3.25  & 7.0  &  14.5  &  -8.8  &  -1.1\\
-0.0117 & 2.7 & 0.0  &  -1.0  &  2.7 - 3.25  & 7.0  &  12.68  &  -8.8  &  -0.6\\
-0.0145 & 2.7 & 0.55  & 0.55  &  2.7 - 3.3  &  5.2  &  13.02  &  -23.2  &  -0.6\\
-0.0145 & 3.0 & 0.55  & 0.55  &  2.7 - 3.35  & 6.2  &  14.85  &  -22.8  &  -1.1\\
-0.0145 & 3.0 & 0.0  & 0.0  &  2.7 - 3.3  &  6.9  &  15.58  &  -22.8  &  -0.6\\
-0.0145 & 2.7 & 0.0  & 0.0  &  2.7 - 3.3  &  6.0  &  14.0  &  -22.8  &  -0.6\\
-0.0145 & 2.7 & 0.0  & 0.55  &  2.7 - 3.3  &  4.2  &  13.22  &  -22.9  &  -1.1\\
-0.0145 & 3.0 & 0.0 & 0.55  &  2.65 - 3.3  &  4.2  &  14.48  &  -22.6  &  -0.6\\
-0.0145 & 3.0 & 0.55  & 0.0  &  2.7 - 3.35  & 6.0  &  15.95 &  -20.8  &  -1.1\\
-0.0145 & 2.7 & 0.55  & 0.0  &  2.7 - 3.3  & 5.0  &  14.18  &  -22.0  &  -0.6\\
\hline
\end{tabular}
\label{table:3}
\end{table}
 \par \hspace{12pt}  The maximum percentage changes in the numerical values of  $g_2$, $g_3$ and $g_4$ with respect to their central values, due to variation of a single parameter, as given in  Table 3, were found to be as follows: due to variation of  $\langle \bar{q}q\rangle$  alone $g_2$, $g_3$ and $g_4$ change by 32\%, 18\% and 78\% respectively; changes in $\alpha$ and $\beta$ lead to  47\%, 12\% and 20\% changes in the values of $g_2$, $g_3$ and $g_4$ respectively; finally variation of $s_\pi$ changes $g_2$, $g_3$ and $g_4$ by 18\%, 13\% and 10\% respectively.  Looking at the contributing components to B$[\Pi_{+o}^{ij}]_{OPE}$, we find that the continuum contribution for the diagonal cases with (ij)=(NN), $(N_+N_+)$ and $(N_-N_-)$ are upto 46\% of the total sum of the respective case and the last term in the OPE contribution is upto  5\% of the same. For the non-diagonal cases the situation is not so favorable. Continuum contribution to  $B[\Pi_{+o}^{N_-N}]_{OPE}$ is (1.5 - 2.0) times larger than the sum, considered as a limiting case for the validity of the sum rule \cite{ioffe2010}, and the last term   included in the OPE is upto 1/3  of the sum. For the   $B[\Pi_{+o}^{N_+N}]_{OPE}$, the situation is worse than this. However, we may look at this problem from a different perspective. Allowing for the mixing of the states with different energies but with the same quantum numbers, calling $u_+$=$(u_{N_+}+u_N)/2$, $u_-$=$(u_{N_+}-u_N)/2$  and  $\Pi_{+o}^{(1)}$=$\bar{u}_+(q)\gamma_0 \Pi(q,p) \gamma_0 u_N(q-p)$, $\Pi_{+o}^{(2)}$=$\bar{u}_-(q)\gamma_0 \Pi(q,p) \gamma_0 u_N(q-p)$, we find that the largest  continuum contribution in the considered range of $M^2$ is for B[$\Pi_{+o}^{(2)}$] and is 82\% of the total sum whereas the largest contribution for the last term in the OPE is for $\Pi_{+o}^{(1)}$ and is 13\% of the total sum. It is in this sense that we consider all the five sum rules as acceptable within the range of $M^2$ considered. We also tried  $\Pi_{+0}^{N_-N_+}$ sum rule and found that this utterly fails to do the job.   \\   
\par \hspace{12pt} It was observed that the renormalization group improvement of  $(\Pi_{+o}^{ij})_{OPE}$ through the introduction of L has a significant bearing on the final results of the coupling constants  $g_i$'s. \\ \par \hspace{12pt}  Based on the analysis of our results, we find the following values of the coupling constants of the pion with the nucleon and its resonances :
\begin{equation}
\begin{aligned}
g_2\equiv g_{N_+\pi^0 N_+}=5.95\pm 1.95,  g_3\equiv g_{N_-\pi^0N_-}=13.6\pm 2.35, \\ g_4\equiv g_{N_+\pi^0 N}=-16.0\pm7.2,  g_5\equiv g_{N_-\pi^0N}=-0.85\pm 0.25,  
\end{aligned}
\end{equation}
where we have introduced the conventioinal notation for the sake of clarity and the result for $g_5$ is not our independent result as stated earlier. To the best of our knowledge, the results  on $g_{N_+\pi^0 N_+}$ and $g_{N_+\pi^0N}$ are new in this work. 

\section{Discussions}
 Though in this approach of projected correlation function one takes into account all the Lorentz structures  simultaneously, due to appearance of  $E_0(s_{\pi}/M^2)$ in Eqs.(14) and (15), the model continuum contribution is kept low for a given sum rule. By first taking different matrix elements between a pair of states from among nucleon and nucleon resonances and then applying dispersion relation, we change the residues of the poles while the arguments of the coupling constants $g_i(q_0)$ are decided by the pole position only. We exploited this fact to get different algebraic equations which were solved numerically for the desired coupling constants. We have applied this approach of  projected correlation function in a case where multiple of states are simultaneously and explicitly taken into account. The appearance of derivative of the coupling constant, $g'_i$, is intrinsic to this approach. We have made a reasonable assumption that $g'_i/g_i$ for i=2, 3, 4, 5 is approximately the same as $g'_1/g_1$, though they all may not be the same at a time and the latter can be determined separately.\\

 \par \hspace{12pt} In Ref.\cite{aliev} the authors have applied light-cone QCD sum rule with all possible Lorentz structures in the correlation function and their derivatives. In addition, they have used  $s_\pi$= (4.0, 4.5) Ge$V^2$ and $1.5 GeV^2 <M^2 <2.5 GeV^2$ and they get $g_{N_-\pi^0N_-}$=$10\pm2$. In Ref. \cite{azizi} authors have used QCD sum rule based on three-point function and have obtained $g_{N_-\pi^0N_-}$=$6.56\pm 1.96$, $g_{N_-\pi^0 N}$=$0.882\pm0.264$. An and Saghai \cite{an11} have applied chiral constituent quark approach considering $N_-$-resonance as a mixture of a three-quark state with exact spin-flavor and orbital  $SU(6)\times O(3)$ symmetry, and a five quark state; the authors were able to get satisfactory results with respect  to all known partial decay widths of $N_-$-resonace with  $g_{N_-\pi^0 N}\simeq -0.6$ in particle basis and  $g_{N_-\pi^0 N}\simeq -1.1$ in the isospin basis. Treating baryons in a mirror assignment where $N_-$ is considered predominantly as the chiral partner of the nucleon, Olbrich et al. \cite{olbrich18} have got $g_{N_-\pi N}\simeq-0.7$. Jido et al \cite{jido98} have pointed out that in the soft pion limit, $g_{ N_-\pi N}$ coupling constant vanishes due to chiral symmetry while    $g_{ N_-\eta N}$ coupling constant remains finite.\\
\par \hspace{12pt} Our spectral function does not include N(1650)-state which has the same quantum numbers as N(1535)-state and is in so close proximity to it that it may have a significant bearing on the determination of the coupling constants $g_i$'s. The widths of both of these states are $\sim$100 MeV and our guess is that taking N(1650)-state into account will lead to decrease in the numerical values of the coupling constants $g_i$'s.
\begin{figure}[h]
\centering
\includegraphics[width=0.45\linewidth]{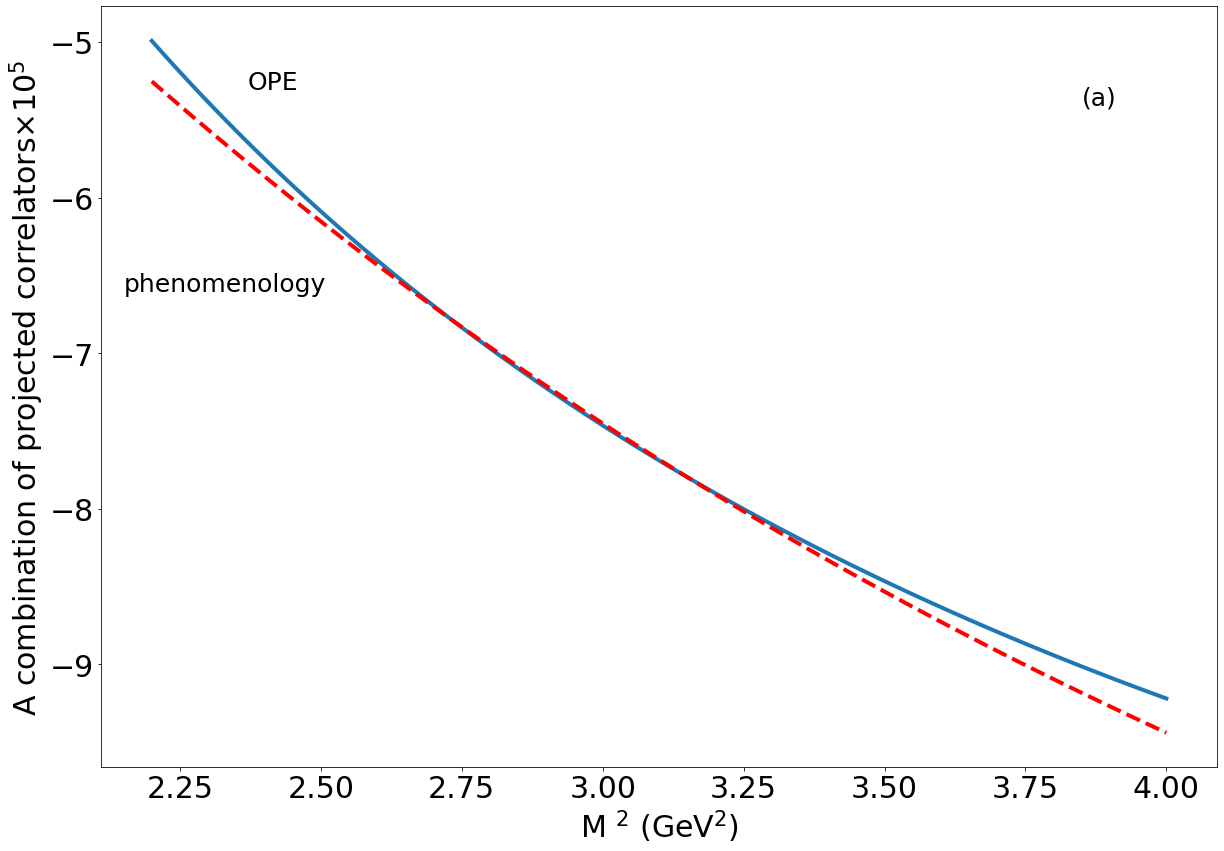}
\includegraphics[width=0.45\linewidth]{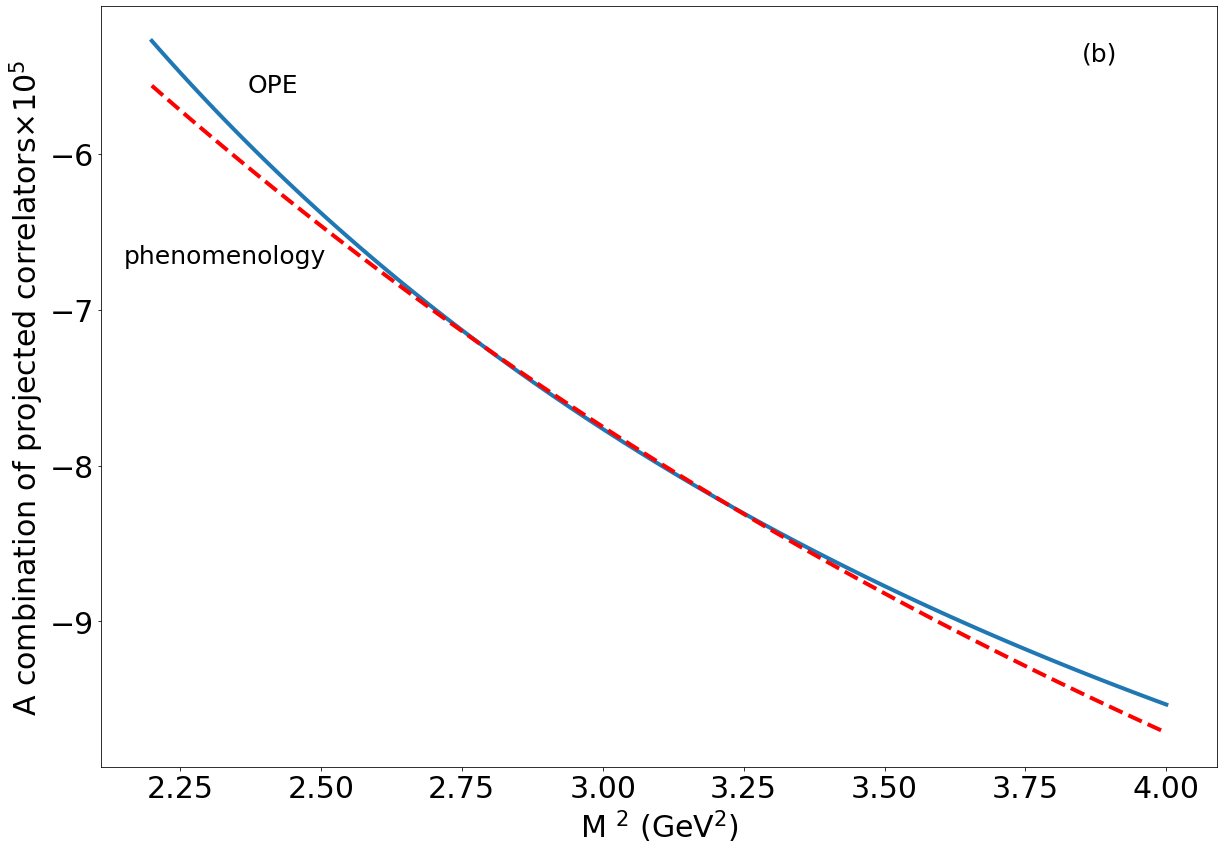}
\caption{{\scriptsize A combination of projected correlation functions plotted as a function of  Borel mass squared, $M^2$.  The values of parameters used for Fig. (a) are : $\alpha=-1.0$, $\beta=-1.0$, $s_\pi$=2.7 Ge$V^2$, $\langle \bar{q}q\rangle$=$-0.0117 GeV^3$, and $g_2$=6.0,  $g_3$=13.0, $g_4=-10.5$, $g_5=-1.1$; for Fig. (b)  the values of the parameters are : $\alpha$=0.0, $\beta$=0.0, $s_\pi$=2.7 Ge$V^2$, $\langle \bar{q}q\rangle$=$-0.0117 GeV^3$, and $g_2$=7.9, $g_3$=11.48, $g_4=-10.4$, $g_5=-0.6$}}
\label{fig:fig2}
\vspace{38pt}
\includegraphics[width=0.45\linewidth]{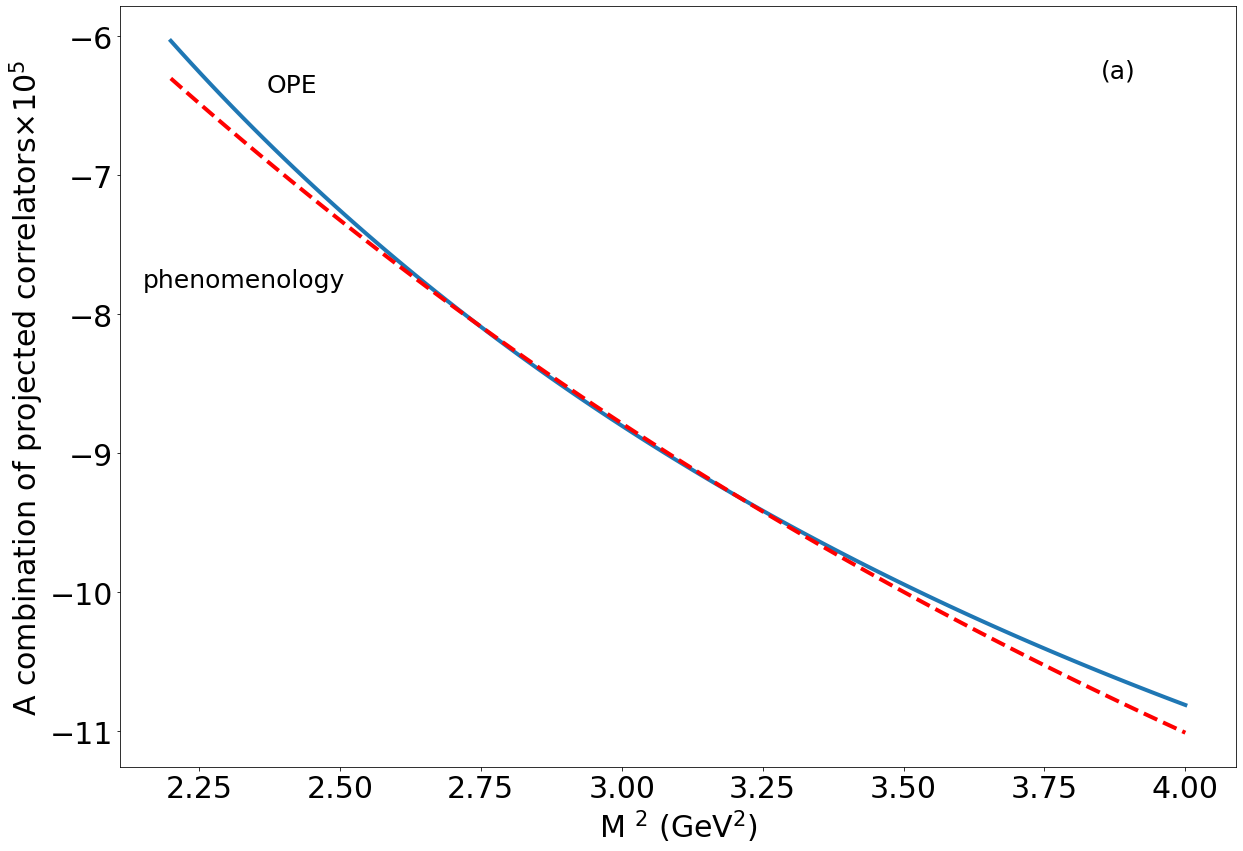}
\includegraphics[width=0.45\linewidth]{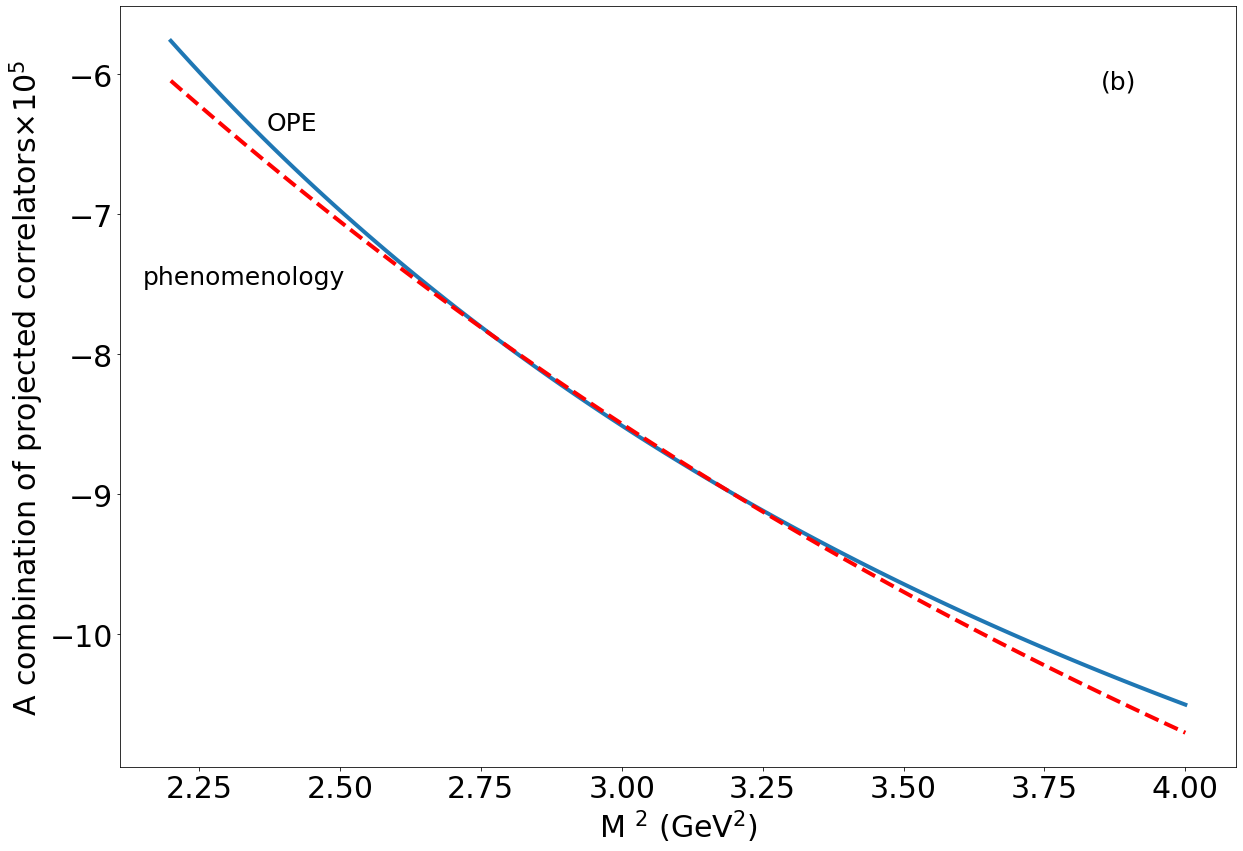}
\caption{{\scriptsize A combination of projected correlation functions plotted as a function of  Borel mass squared, $M^2$.  The values of parameters used for Fig. (a)  are : $\alpha=-1.0$, $\beta$=0.0, $s_\pi$=3.0 Ge$V^2$, $\langle \bar{q}q\rangle$=$-0.0117 GeV^3$, and $g_2$=7.0, $g_3$=12.66, $g_4=-11.2$, $g_5=-0.6$; for Fig. (b)  the values of the parameters are : $\alpha$=0.0, $\beta=-1.0$, $s_\pi$=3.0 Ge$V^2$, $\langle \bar{q}q\rangle$=$-0.0117 GeV^3$, and $g_2$=7.0, $g_3$=14.5, $g_4=-8.8$,$g_5=-1.1$}}
\label{fig:fig3}
\end{figure}

\pagebreak
\newpage
\begin{figure}[h]
\centering
\includegraphics[width=0.45\linewidth]{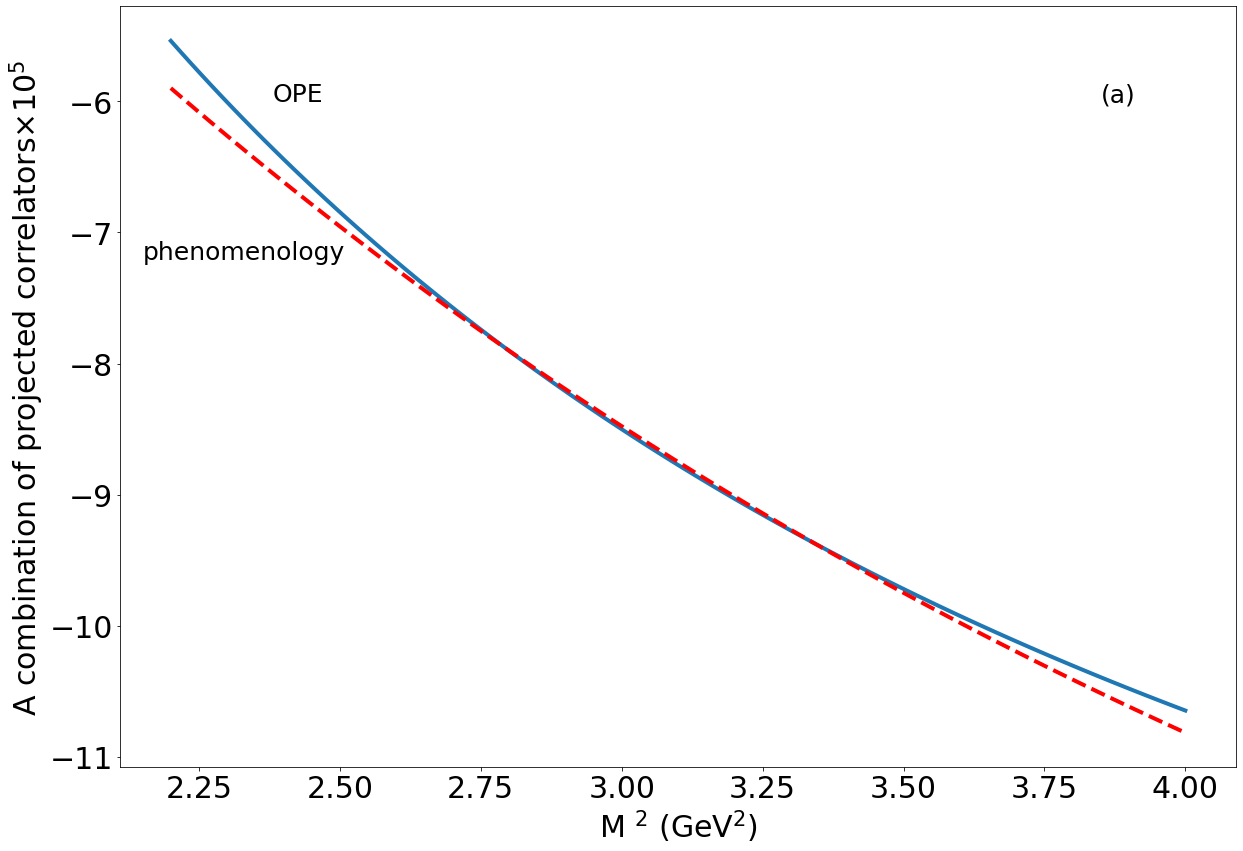}
\includegraphics[width=0.45\linewidth]{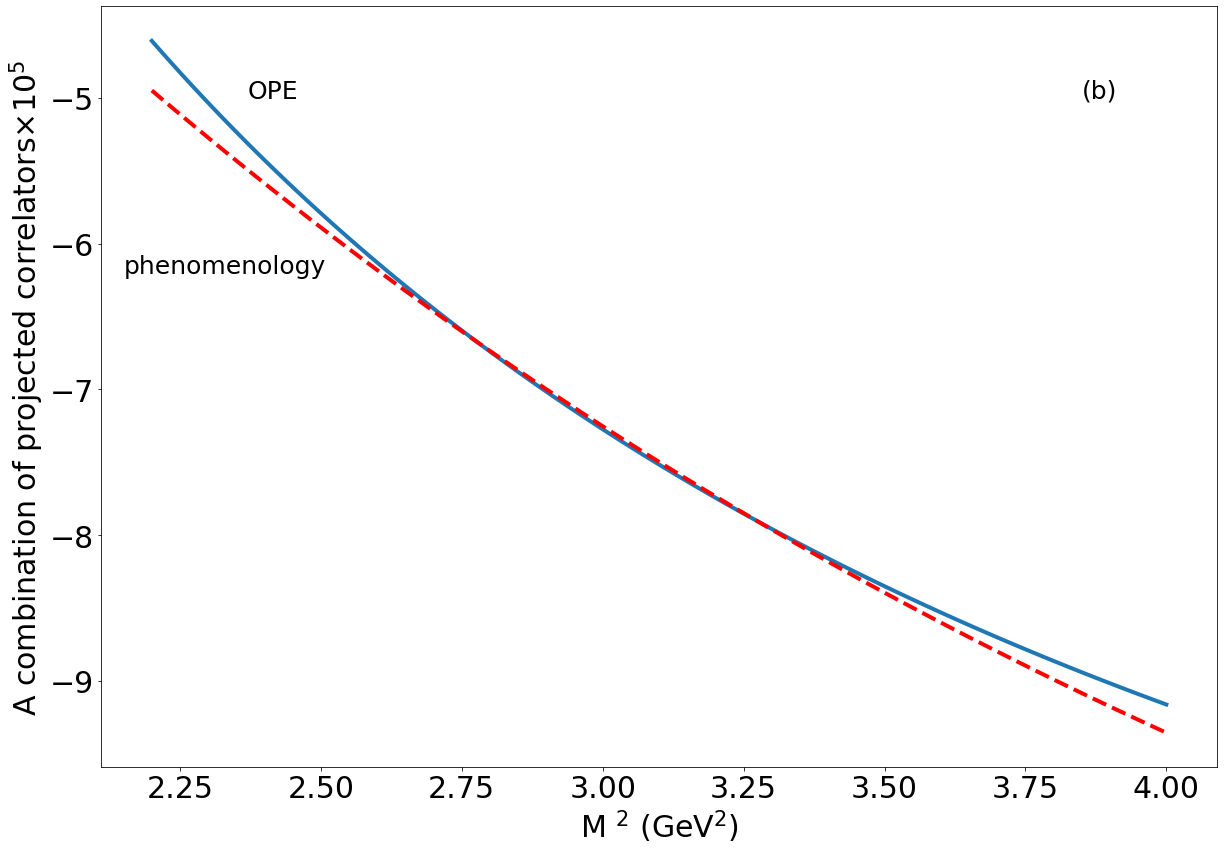}
\caption{{\scriptsize A combination of projected correlation functions plotted as a function of  Borel mass squared, $M^2$.  The values of parameters used for Fig. (a) are : $\alpha$=0.55, $\beta$=0.55, $s_\pi$=3.0 Ge$V^2$, $\langle \bar{q}q\rangle$=$-0.0145 GeV^3$, and $g_2$=6.2, $g_3$=14.85, $g_4=-22.8$, $g_5=-1.1$; for Fig. (b)  the values of the parameters are : $\alpha$=0.0, $\beta$=0.0, $s_\pi$=2.7 Ge$V^2$, $\langle \bar{q}q\rangle$=$-0.0145 GeV^3$, and $g_2$=6.0, $g_3$=14.0, $g_4=-22.8$, ,$g_5=-0.6$}}
\label{fig:fig4}
\vspace{38pt}

\includegraphics[width=0.45\linewidth]{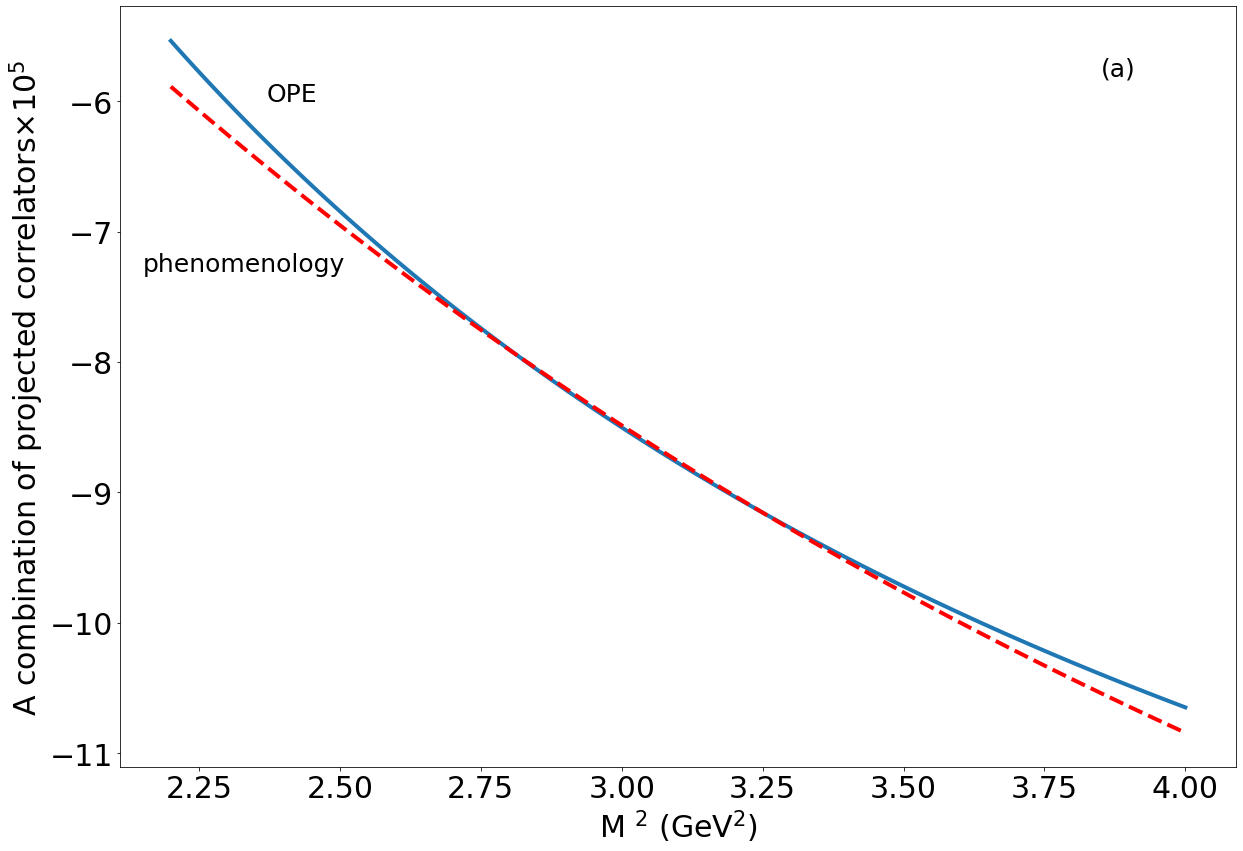}
\includegraphics[width=0.45\linewidth]{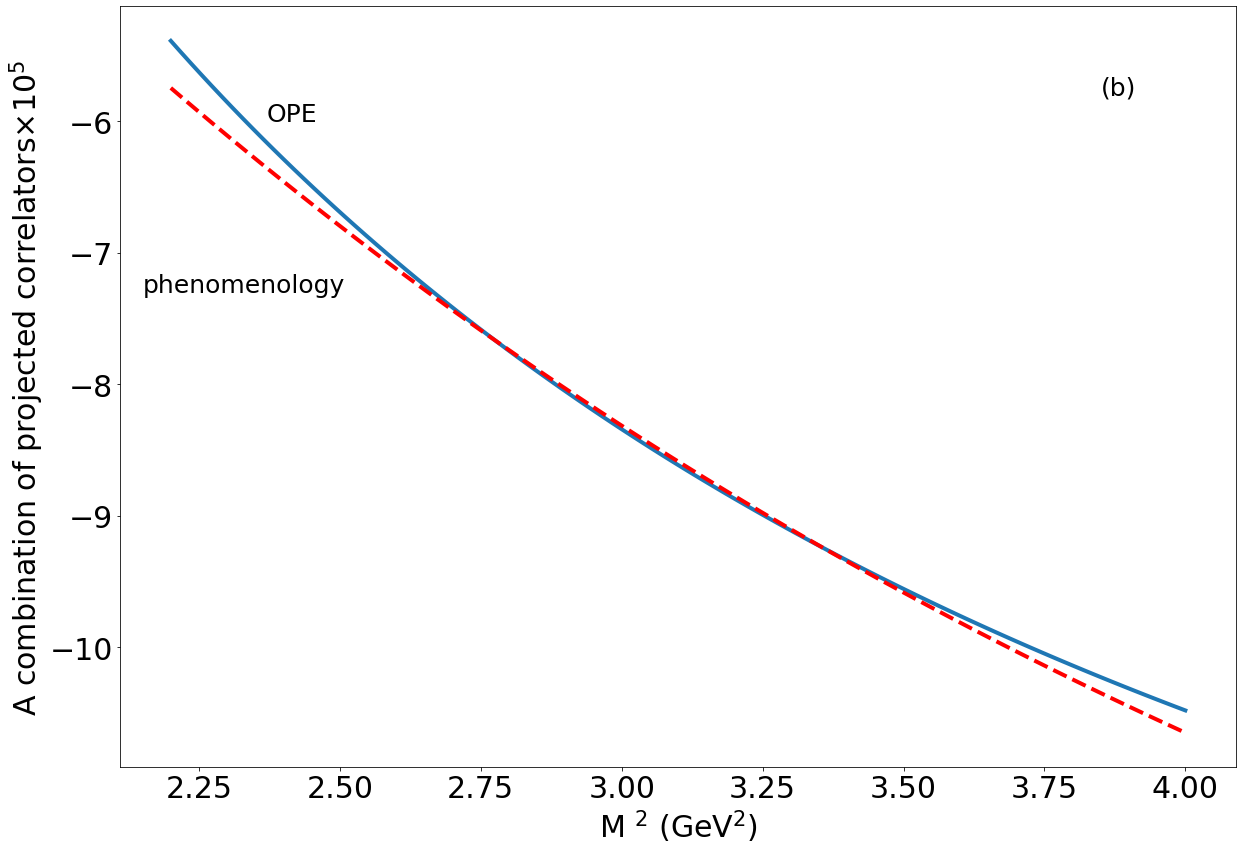}
\caption{{\scriptsize A combination of projected correlation functions plotted as a function of  Borel mass squared, $M^2$.  The values of parameters used for Fig. (a) are : $\alpha$=0.0, $\beta$=0.55, $s_\pi$=3.0 Ge$V^2$, $\langle \bar{q}q\rangle$=$-0.0145GeV^3$, and $g_2$=4.2, $g_3$=14.48, $g_4=-22.6$, $g_5=-0.6$; for Fig. (b)  the values of the parameters are : $\alpha$=0.55, $\beta$=0.0, $s_\pi$=3.0 Ge$V^2$, $\langle \bar{q}q\rangle$=$-0.0145 GeV^3$, and $g_2$=6.0, $g_3$=15.95, $g_4=-20.8$, $g_5=-1.1$}}
\label{fig:fig5}
\end{figure}
\renewcommand{\theequation}{A-\arabic{equation}}
\setcounter{equation}{0}
\section*{APPENDIX}
Using the phenomenological  Lagrangian $\mathcal{L}$ as given in Eq. (6), we can calculate the correlation function and this gives
\begin{equation}
\begin{aligned}
\Pi (q,p)_{ph}=g_1 \lambda^2 \frac{(\not{q}+m)i\gamma_5(\not{q}-\not{p}+m)}{(q^2-m^2)[(q-p)^2-m^2]}+g_2 \lambda_1^2 \frac{(\not{q}+m_1)i\gamma_5(\not{q}-\not{p}+m_1)}{(q^2-m_1^2)[(q-p)^2-m_1^2]}\\+g_3 \lambda_2^2 \frac{\gamma_5(\not{q}+m_2)i\gamma_5(\not{q}-\not{p}+m_2)(-\gamma_5)}{(q^2-m_2^2)[(q-p)^2-m_2^2]}\\
+g_4\lambda \lambda_1\Big[ \frac{(\not{q}+m_1)i\gamma_5(\not{q}-\not{p}+m)}{(q^2-m_1^2)[(q-p)^2-m^2]}+\frac{(\not{q}+m)i\gamma_5(\not{q}-\not{p}+m_1)}{(q^2-m^2)[(q-p)^2-m_1^2]}\Big]\\
+g_5\lambda \lambda_2\Big[ \frac{i\gamma_5(\not{q}+m_2)(\not{q}-\not{p}+m)}{(q^2-m_2^2)[(q-p)^2-m^2]}+\frac{(\not{q}+m)(\not{q}-\not{p}+m_2)i\gamma_5}{(q^2-m^2)[(q-p)^2-m_2^2]}\Big]\\
+g_6\lambda_1 \lambda_2\Big[ \frac{i\gamma_5(\not{q}+m_2)(\not{q}-\not{p}+m_1)}{(q^2-m_2^2)[(q-p)^2-m_1^2]}+\frac{(\not{q}+m_1)(\not{q}-\not{p}+m_2)i\gamma_5}{(q^2-m_1^2)[(q-p)^2-m_2^2]}\Big]\\+\text{contribution  from   continuum   and   higher   states}
\end{aligned}
\end{equation}
\par \hspace{20pt}   Below we list vacuum-to-pion matrix elements of non-local operators used in the text based on Refs.\cite{belyaev, doi04}:
\begin{equation}\langle 0| \bar{u}(0)i\gamma_5 u(x)  |\pi^0(p)\rangle    =-\frac{\langle\bar{u}u\rangle }{f_\pi} +\textit{ip.x}\frac{\langle\bar{u}u\rangle }{2f_\pi} +(\textit{p.x})^2\Big[ \frac{\langle\bar{u}u\rangle }{6 f_\pi}-\frac{18.91}{2\sqrt{2}}f_{3\pi}\Big] \end{equation} 
\begin{equation}
\begin{aligned}
\langle 0| \bar{u}(0)\gamma_\mu \gamma_5 u(x)  |\pi^0(p)\rangle   = i f_\pi p_\mu( 1-\frac{i}{2} \textit{p.x}-\frac{(\textit{p.x})^2}{2} a') -\frac{i}{18}f_\pi \delta^2(\textit{p.x} x_\mu \\ -\frac{5}{2} x^2 p_\mu) (1-\frac{i}{2} \textit{p.x}) 
\end{aligned}
\end{equation}
\begin{equation}\langle 0| \bar{u}(0)\gamma_5 \sigma^{\mu \nu} u(x)  |\pi^0(p)\rangle    =-i(p_{\mu} x_{\nu}-p_{\nu} x_{\mu})\frac{\langle\bar{u}u\rangle }{6f_{\pi}}\Big[1-\frac{i}{2}\textit{p.x}-\frac{c'}{2}(p.x)^2\Big]\end{equation}
\begin{equation}
\begin{aligned}
\langle 0|u^a(x) g_s G^n_{\mu \nu} (x/2)\bar{u}^b(0)|\pi^0(p)\rangle  =i\frac{f_{3\pi}}{16\sqrt{2}} t^n_{ab}\gamma_5 (\sigma_{\lambda \mu}p_\nu-\sigma_{\lambda \nu} p_\mu)p^\lambda\\ +\frac{if_\pi \delta^2}{252.16}t^n_{ab}\gamma_5 [10(\gamma_\mu p_\nu-\gamma_\nu p_\mu) \textit{p.x}+16 \not{p}(p_\mu x_\nu- p_\nu x_\mu)]\\ -\frac{i}{48}t^n_{ab} f_\pi \delta^2 \epsilon_{\mu\nu}^{\quad  \alpha\beta}\gamma_\alpha p_\beta (1-\frac{i}{2} \textit{p.x})
\end{aligned}
\end{equation}
\par \hspace{12pt}   Following is the Borel transform of a function used in the text \cite{ioffe2010, kondo03}:
\begin{equation} B [f( p_0^2)]=\lim\limits_{\substack{
                                          n \to \infty\\
                                         -p_0^2\to\infty\\
                                         -p_0^2/n=M^2
                                         }}
                                         \frac{(-p_0^2)^{n+1}}{n!}\Big(\frac{d}{dp_0^2}\Big)^n f(p_0^2) \end{equation}
\bibliographystyle{unsrt}
\bibliography{Docr7}
\end{document}